\documentstyle[twoside,psfig]{article}

\catcode`\@=11
\long\def\@makefntext#1{
\protect\noindent \hbox to 3.2pt {\hskip-.9pt  
$^{{\eightrm\@thefnmark}}$\hfil}#1\hfill}		

\def\thefootnote{\fnsymbol{footnote}}
\def\@makefnmark{\hbox to 0pt{$^{\@thefnmark}$\hss}}	
	
\def\ps@myheadings{\let\@mkboth\@gobbletwo
\def\@oddhead{\hbox{}
\rightmark\hfil\eightrm\thepage}   
\def\@oddfoot{}\def\@evenhead{\eightrm\thepage\hfil
\leftmark\hbox{}}\def\@evenfoot{}
\def\sectionmark##1{}\def\subsectionmark##1{}}

\oddsidemargin=\evensidemargin
\renewcommand{\thefootnote}{\fnsymbol{footnote}}

\newcounter{sectionc}\newcounter{subsectionc}\newcounter{subsubsectionc}
\renewcommand{\section}[1] {\vspace{12pt}\addtocounter{sectionc}{1} 
\setcounter{subsectionc}{0}\setcounter{subsubsectionc}{0}\noindent 
	{\tenbf\thesectionc. #1}\par\vspace{5pt}}
\renewcommand{\subsection}[1] {\vspace{12pt}\addtocounter{subsectionc}{1} 
	\setcounter{subsubsectionc}{0}\noindent 
	{\bf\thesectionc.\thesubsectionc. {\kern1pt \bfit #1}}\par\vspace{5pt}}
\renewcommand{\subsubsection}[1] {\vspace{12pt}\addtocounter{subsubsectionc}{1}
	\noindent{\tenrm\thesectionc.\thesubsectionc.\thesubsubsectionc.
	{\kern1pt \tenit #1}}\par\vspace{5pt}}
\newcommand{\nonumsection}[1] {\vspace{12pt}\noindent{\tenbf #1}
	\par\vspace{5pt}}

\newcounter{appendixc}
\newcounter{subappendixc}[appendixc]
\newcounter{subsubappendixc}[subappendixc]
\renewcommand{\thesubappendixc}{\Alph{appendixc}.\arabic{subappendixc}}
\renewcommand{\thesubsubappendixc}
	{\Alph{appendixc}.\arabic{subappendixc}.\arabic{subsubappendixc}}

\renewcommand{\appendix}[1] {\vspace{12pt}
        \refstepcounter{appendixc}
        \setcounter{figure}{0}
        \setcounter{table}{0}
        \setcounter{lemma}{0}
        \setcounter{theorem}{0}
        \setcounter{corollary}{0}
        \setcounter{definition}{0}
        \setcounter{equation}{0}
        \renewcommand{\thefigure}{\Alph{appendixc}.\arabic{figure}}
        \renewcommand{\thetable}{\Alph{appendixc}.\arabic{table}}
        \renewcommand{\theappendixc}{\Alph{appendixc}}
        \renewcommand{\thelemma}{\Alph{appendixc}.\arabic{lemma}}
        \renewcommand{\thetheorem}{\Alph{appendixc}.\arabic{theorem}}
        \renewcommand{\thedefinition}{\Alph{appendixc}.\arabic{definition}}
        \renewcommand{\thecorollary}{\Alph{appendixc}.\arabic{corollary}}
        \renewcommand{\theequation}{\Alph{appendixc}.\arabic{equation}}
        \noindent{\tenbf Appendix \theappendixc #1}\par\vspace{5pt}}
\newcommand{\subappendix}[1] {\vspace{12pt}
        \refstepcounter{subappendixc}
        \noindent{\bf Appendix \thesubappendixc. {\kern1pt \bfit #1}}
	\par\vspace{5pt}}
\newcommand{\subsubappendix}[1] {\vspace{12pt}
        \refstepcounter{subsubappendixc}
        \noindent{\rm Appendix \thesubsubappendixc. {\kern1pt \tenit #1}}
	\par\vspace{5pt}}

\newcommand{\textlineskip}{\baselineskip=13pt}
\newcommand{\smalllineskip}{\baselineskip=10pt}

\def\eightcirc{
\begin{picture}(0,0)
\put(4.4,1.8){\circle{6.5}}
\end{picture}}
\def\eightcopyright{\eightcirc\kern2.7pt\hbox{\eightrm c}}

\def\abstracts#1#2#3{{
	\centering{\begin{minipage}{4.5in}\baselineskip=10pt\footnotesize
	\parindent=0pt #1\par 
	\parindent=15pt #2\par
	\parindent=15pt #3
	\end{minipage}}\par}} 

\renewenvironment{thebibliography}[1]
	{\frenchspacing
	 \ninerm\baselineskip=11pt
	 \begin{list}{\arabic{enumi}.}
	{\usecounter{enumi}\setlength{\parsep}{0pt}
	 \setlength{\leftmargin 12.7pt}{\rightmargin 0pt} 
	 \setlength{\itemsep}{0pt} \settowidth
	{\labelwidth}{#1.}\sloppy}}{\end{list}}

\newcounter{itemlistc}
\newcounter{romanlistc}
\newcounter{alphlistc}
\newcounter{arabiclistc}

\newcommand{\fcaption}[1]{
        \refstepcounter{figure}
        \setbox\@tempboxa = \hbox{\footnotesize Fig.~\thefigure. #1}
        \ifdim \wd\@tempboxa > 5in
           {\begin{center}
        \parbox{5in}{\footnotesize\smalllineskip Fig.~\thefigure. #1}
            \end{center}}
        \else
             {\begin{center}
             {\footnotesize Fig.~\thefigure. #1}
              \end{center}}
        \fi}

\newcommand{\tcaption}[1]{
        \refstepcounter{table}
        \setbox\@tempboxa = \hbox{\footnotesize Table~\thetable. #1}
        \ifdim \wd\@tempboxa > 5in
           {\begin{center}
        \parbox{5in}{\footnotesize\smalllineskip Table~\thetable. #1}
            \end{center}}
        \else
             {\begin{center}
             {\footnotesize Table~\thetable. #1}
              \end{center}}
        \fi}

\def\@citex[#1]#2{\if@filesw\immediate\write\@auxout
	{\string\citation{#2}}\fi
\def\@citea{}\@cite{\@for\@citeb:=#2\do
	{\@citea\def\@citea{,}\@ifundefined
	{b@\@citeb}{{\bf ?}\@warning
	{Citation `\@citeb' on page \thepage \space undefined}}
	{\csname b@\@citeb\endcsname}}}{#1}}

\newif\if@cghi
\def\cite{\@cghitrue\@ifnextchar [{\@tempswatrue
	\@citex}{\@tempswafalse\@citex[]}}
\def\citelow{\@cghifalse\@ifnextchar [{\@tempswatrue
	\@citex}{\@tempswafalse\@citex[]}}
\def\@cite#1#2{{$\null^{#1}$\if@tempswa\typeout
	{IJCGA warning: optional citation argument 
	ignored: `#2'} \fi}}

\def\pmb#1{\setbox0=\hbox{#1}
	\kern-.025em\copy0\kern-\wd0
	\kern.05em\copy0\kern-\wd0
	\kern-.025em\raise.0433em\box0}

\def\fnm#1{$^{\mbox{\scriptsize #1}}$}
\def\fnt#1#2{\footnotetext{\kern-.3em
	{$^{\mbox{\scriptsize #1}}$}{#2}}}

\def\fpage#1{\begingroup
\voffset=.3in
\thispagestyle{empty}\begin{table}[b]\centerline{\footnotesize #1}
	\end{table}\endgroup}

\def\runninghead#1#2{\pagestyle{myheadings}
\markboth{{\protect\footnotesize\it{\quad #1}}\hfill}
{\hfill{\protect\footnotesize\it{#2\quad}}}}
\font\tenrm=cmr10
\font\tenit=cmti10 
\font\tenbf=cmbx10
\font\bfit=cmbxti10 at 10pt
\font\ninerm=cmr9

\font\eightrm=cmr8

\def\qed{\hbox{${\vcenter{\vbox{			
   \hrule height 0.4pt\hbox{\vrule width 0.4pt height 6pt
   \kern5pt\vrule width 0.4pt}\hrule height 0.4pt}}}$}}

\renewcommand{\thefootnote}{\fnsymbol{footnote}}	

\def\Del{\Delta}
\def\dof{{\rm d.o.f.}}
\def\xa{x_\alpha^{}}
\def\contact{g_E^2/c^2_\chi m_{Z_2}^2}
\def\xibar{\bar{\xi}}
\def\tnew{T_{\rm new}}
\def\to{\rightarrow}
\def\ie{{\it i.e.}}
\def\eg{{\it e.g.}}
\def\vsk#1{\noalign{\vskip#1 cm}}
\def\vsp#1{\vspace{#1 cm}}
\def\hsp#1{\hspace{#1 cm}}
\def\ov{\overline}
\def\disp{\displaystyle}
\def\gsim{~{\rlap{\lower 3.5pt\hbox{$\mathchar\sim$}}\raise 1pt\hbox{$>$}}\,}
\def\lsim{~{\rlap{\lower 3.5pt\hbox{$\mathchar\sim$}}\raise 1pt\hbox{$<$}}\,}
\def\hph{\hphantom{-}}
\def\hpz{\hphantom{0}}
\def\fb{{\rm FB}}
\def\alps{\alpha_s}
\def\gev{~{\rm GeV}}
\def\tev{~{\rm TeV}}
\def\mt{m_t^{}}
\def\mh{m_H^{}}
\def\xh{x_H^{}}
\def\mw{m_W^{}}
\def\ds{\Del S}
\def\dt{\Del T}
\def\du{\Del U}
\def\sbar{\bar{s}^2}
\def\gzbar{\bar{g}_Z^2}
\def\gwbar{\bar{g}_W^2}
\def\abar{\bar{\alpha}}
\def\dgzbar{\Del \gzbar}
\def\dsbar{\Del \sbar}
\def\etal{{\it et al.~}}
\newcommand{\beq}{\begin{equation}}
\newcommand{\eeq}{\end{equation}}
\newcommand{\bea}{\begin{eqnarray}}
\newcommand{\eea}{\end{eqnarray}}
\newcommand{\bsub}{\begin{subequations}}
\newcommand{\esub}{\end{subequations}}
\def\MPLA#1#2#3{Mod. Phys. Lett. {\bf A#1} (#2) #3}
\def\PRD#1#2#3{Phys. Rev. {\bf D#1} (#2) #3}
\def\NPB#1#2#3{Nucl. Phys. {\bf B#1} (#2) #3}
\def\PTP#1#2#3{Prog. Theor. Phys. {\bf #1} (#2) #3}
\def\ZPC#1#2#3{Z. Phys. {\bf C#1} (#2) #3}
\def\EPJC#1#2#3{Eur. Phys. J. {\bf C#1} (#2) #3}
\def\PLB#1#2#3{Phys. Lett. {\bf B#1} (#2) #3}
\def\PRL#1#2#3{Phys. Rev. Lett. {\bf #1} (#2) #3}
\def\PR#1#2#3{Phys. Rep. {\bf #1} (#2) #3}
\makeatletter
\newtoks\@stequation

\def\subequations{\refstepcounter{equation}%
  \edef\@savedequation{\the\c@equation}%
  \@stequation=\expandafter{\theequation}
  \edef\@savedtheequation{\the\@stequation}
  \edef\oldtheequation{\theequation}%
  \setcounter{equation}{0}%
  \def\theequation{\oldtheequation\alph{equation}}}

\def\endsubequations{%
  \ifnum\c@equation < 2 \@warning{Only \the\c@equation\space subequation
    used in equation \@savedequation}\fi
  \setcounter{equation}{\@savedequation}%
  \@stequation=\expandafter{\@savedtheequation}%
  \edef\theequation{\the\@stequation}%
  \global\@ignoretrue}

\def\eqnarray{\stepcounter{equation}\let\@currentlabel\theequation
\global\@eqnswtrue\m@th
\global\@eqcnt\z@\tabskip\@centering\let\\\@eqncr
$$\halign to\displaywidth\bgroup\@eqnsel\hskip\@centering
     $\displaystyle\tabskip\z@{##}$&\global\@eqcnt\@ne
      \hfil$\;{##}\;$\hfil
     &\global\@eqcnt\tw@ $\displaystyle\tabskip\z@{##}$\hfil
   \tabskip\@centering&\llap{##}\tabskip\z@\cr}

\makeatother
\begin{document}

\runninghead{
Looking for $Z'$ bosons in Supersymmetric $E_6$ Models 
through Electroweak Precision Data}
{Looking for $Z'$ bosons in Supersymmetric $E_6$ Models 
through Electroweak Precision Data}

\normalsize\textlineskip
\thispagestyle{empty}
\setcounter{page}{1}


\vspace*{-2.5cm}
\begin{flushright}
\smalllineskip{
{\small\bf 
	SNS-PH/00-03\\
	hep-ph/0002128
}}
\end{flushright}

\vspace*{0.88truein}

\fpage{1}
\centerline{\bf 
LOOKING FOR $Z'$ BOSONS IN SUPERSYMMETRIC $E_6$ MODELS}
\vspace*{0.035truein}
\centerline{\bf THROUGH ELECTROWEAK PRECISION DATA}
\vspace*{0.37truein}
\centerline{\footnotesize GI-CHOL CHO}
\vspace*{0.015truein}
\centerline{\footnotesize\it 
Scuola Normale Superiore,} 
\baselineskip=10pt
\centerline{\footnotesize\it 
Piazza dei Cavalieri 7, Pisa 56126, Italy}
\vspace*{10pt}

\vspace*{0.21truein}
\abstracts{
We review constraints on additional $Z'$ bosons predicted 
in supersymmetric (SUSY) $E_6$ models from electroweak experiments 
-- $Z$-pole experiments, $\mw$ measurements and the low-energy 
neutral current (LENC) experiments. 
Four representative models -- $\chi,\psi,\eta, \nu$ models -- 
are studied in some detail. 
We find that the improved data of parity violation in cesium atom, 
which is 2.2-$\sigma$ away from the Standard Model (SM) prediction, 
could be explained by the exchange of the heavy mass 
eigenstate $Z_2$ in the intermediate state. 
The improvement over the SM can be found in $\chi, \eta, \nu$ 
models, where the total $\chi^2$ of the fit to the 26 data points 
decreases by about five units, owing to the better fit to 
the atomic parity violation. 
Impacts of the kinetic mixing between the U(1)$_Y$ and U(1)$'$ 
gauge bosons on the $\chi^2$-analysis are studied. 
We find that the $Z'$ model with $(\beta_E, \delta)=(-\pi/4,0.2)$, 
where $\beta_E$ is the mixing angle between $Z_\chi$ and $Z_\psi$ 
bosons and $\delta$ denotes the kinetic mixing,  
shows the most excellent fit to the data: the total $\chi^2$ 
decreases by about seven units as compared to the SM. 
We introduce the effective mixing parameter $\zeta$, a combination 
of the mass and the kinetic mixing parameters. 
The 95\% CL lower mass bound of $Z_2$ can be shown as a 
function of $\zeta$. 
A theoretical prediction on $\zeta$ and the U(1)$'$ gauge coupling 
$g_E$ is studied for the $\chi,\psi,\eta$ and $\nu$ models by  
assuming the minimal particle content of the SUSY $E_6$ models. 
}{}{}
%
%
%
%
\vspace*{1pt}\textlineskip	
\section{Introduction}	
\vspace*{-0.5pt}
\noindent
The presence of an additional $Z'$ boson is predicted in a certain 
class of grand unified theories (GUT) with a gauge group whose 
rank is higher than that of the Standard Model (SM). 
The supersymmetric (SUSY) $E_6$ models are the promising candidates 
which predict the additional $Z'$-boson at the weak 
scale~\cite{hewett_rizzo}. 
Because $E_6$ is a rank-six group, it can have two extra 
${\rm U(1)}$ factors besides the SM gauge group. 
A superposition of the two extra ${\rm U(1)}$ groups may 
survive as the ${\rm U(1)'}$ gauge symmetry at the GUT scale. 
The ${\rm U(1)'}$ symmetry may break spontaneously at the weak 
scale through the radiative corrections to the mass term of the 
SM singlet scalar field~\cite{radiative_u1prime}. 

In general, the additional ${\rm U(1)'}$ gauge boson $Z'$ can mix 
with the hypercharge ${\rm U(1)}_Y$ gauge boson through the kinetic 
term at above the electroweak scale, and also it can mix with the SM 
$Z$ boson after the electroweak symmetry is spontaneously broken.  
Through those mixings, the $Z'$ boson can affect the electroweak 
observables at the $Z$-pole and the $W$-boson mass $\mw$. 
Both the $Z$-$Z'$ mixing and the direct $Z'$ contribution can affect 
neutral current experiments off the $Z$ pole. 
The presence of an additional $Z'$ boson can be explored directly 
at $p \bar{p}$ collider experiments. 

In this review article, we report constraints on $Z'$ bosons in 
the SUSY $E_6$ models from electroweak experiments based on 
the formalism in Refs.~\ref{ref:chu},~\ref{ref:uch}. 
Constraints on the $Z'$ bosons from electroweak experiments have 
been studied by several 
authors~\cite{zprime_old,chm,zprime_lep,chu,erler_langacker}. 
Especially, a special attention has been paid to this 
subject~\cite{apv_zprime} after the new analysis of parity 
violation in cesium atom has led to the improved data of 
the weak charge $Q_W(^{133}_{55}Cs)$~\cite{apv_data},  
which is 2.2-$\sigma$ away from the SM prediction 
(see Table~\ref{tab:ewdata}). 
The analysis given in this article updates their studies by 
allowing for an arbitrary kinetic 
mixing~\cite{holdom,eta_model,general_zzmixing} 
between the $Z'$ boson and the hypercharge $B$ boson. 
The constraints on the $Z'$ bosons can be found by using the results 
of $Z$-pole experiments at LEP1 and SLC, and the $\mw$ measurements 
at Tevatron and LEP2. Also the low-energy neutral current (LENC) 
experiments -- lepton-quark, lepton-lepton scattering 
experiments and atomic parity violation (APV) measurements -- 
constrain the direct exchange of $Z'$ boson. 

It has been found~\cite{chu,uch} that 
the lower mass limit of the heavier mass eigenstate $Z_2$ is obtained 
as a function of the effective $Z$-$Z'$ mixing term $\zeta$, which is 
a combination of the mass and kinetic mixings. 
In principle, $\zeta$ is calculable, together with the gauge 
coupling $g_E$, once the particle spectrum of the $E_6$ model 
is specified. 
We show the theoretical prediction for $\zeta$ and $g_E$ 
in the SUSY $E_6$ models by assuming the minimal particle 
content which satisfies the anomaly free condition and the 
gauge coupling unification\fnm{a}\fnt{a}{Consequence in the case of 
the maximal particle content which preserve the perturbative 
unification of the gauge couplings has been studied 
in Ref.~\ref{ref:rizzo_umeda}.}. 

This paper is organized as follows. 
In the next section, we review the additional $Z'$ 
boson in the SUSY $E_6$ models and the generic feature of 
$Z$-$Z'$ mixing in order to fix our notation. 
We show that the effects of $Z$-$Z'$ mixing and direct 
$Z'$ boson contribution are parametrized 
by the following three terms: 
(i) a tree-level contribution to the $T$ parameter~\cite{peskin_takeuchi}, 
$T_{\rm new}$, 
(ii) the effective $Z$-$Z'$ mass mixing angle $\xibar$ 
and 
(iii) a contact term $\contact$ which appears in the low-energy 
processes. 
In Sec.~3, we collect the data of electroweak experiments which 
will be used in our analysis. 
We also present the theoretical framework to calculate the electroweak 
observables. 
In Sec.~4, we show constraints on the $Z'$ bosons from the 
electroweak data. 
The presence of non-zero kinetic mixing between the 
${\rm U(1)}_Y$ and ${\rm U(1)'}$ gauge bosons 
modifies the couplings between the $Z'$ boson and the SM 
fermions. 
We discuss impacts of the kinetic mixing term 
on the $\chi^2$-analysis. 
The 95\% CL lower mass limit of the heavier mass eigenstate $Z_2$ 
in four representative models -- $\chi,\psi,\eta,\nu$ models -- 
is given as a function of the effective $Z$-$Z'$ mixing parameter 
$\zeta$. 
The $\zeta$-independent constraints from the low-energy 
experiments and those from the direct search experiments 
at Tevatron are also discussed. 
In Sec.~5, 
we find the theoretical prediction for $\zeta$ in $\chi, \psi, \eta, 
\nu$ models by assuming the minimal particle contents.  
Stringent $Z_2$ boson mass bounds are found for most models.
Sec.~6 is devoted to summarize this paper.  

\setcounter{footnote}{0}
\renewcommand{\thefootnote}{\alph{footnote}}

\section{$Z$-$Z'$ mixing in supersymmetric $E_6$ model}
\noindent
\subsection{$Z'$ boson in supersymmetric $E_6$ model}
\noindent
Since the rank of $E_6$ is six, it has two ${\rm U(1)}$ factors 
besides the SM gauge group which arise from the following 
decompositions: 
\bea
	\begin{array}{rl}
	E_6 &\supset {\rm SO(10)} \times {\rm U(1)}_\psi  
\\
	&\supset {\rm SU(5)} \times {\rm U(1)}_\chi \times {\rm U(1)}_\psi. 
	\end{array}
\eea
\label{eq:e6_breaking}
An additional $Z'$ boson in the electroweak scale 
can be parametrized as 
a linear combination of the ${\rm U(1)}_\psi$ gauge boson 
$Z_\psi$ and the ${\rm U(1)}_\chi$ gauge boson 
$Z_\chi$ as~\cite{PDG98} 
\beq
Z' = Z_\chi \cos \beta_E + Z_\psi \sin \beta_E. 
\label{eq:zmixing_e6}
\eeq
In this paper, the following $Z'$ models are studied in some detail: 
\bea
\begin{array}{|c||c|c|c|c|} \hline 
~~~\beta_E~~~ & ~~~0~~~ &~~~\pi/2~~~ & \tan^{-1}(-\sqrt{5/3}) 
& \tan^{-1}(\sqrt{15}) \\ \hline
{\rm model} & \chi      & \psi         & \eta      & \nu      
\\  \hline 
\end{array}
\eea
In the SUSY-$E_6$ models, each generation of the SM quarks 
and leptons is embedded into a {\bf 27} representation. 
In Table~1, we show all the matter fields 
contained in a {\bf 27} and their classification in SO(10) 
and SU(5). 
The ${\rm U(1)'}$ charge assignment on the matter fields 
for each model is also given in the same table. 
The normalization of the ${\rm U(1)'}$ charge follows 
that of the hypercharge. 
\begin{table}[ht] 
\begin{center}
\tcaption{
	The hypercharge $Y$ and the ${\rm U(1)'}$ charge $Q_E$ of 
	all the matter fields in a {\bf 27} for the $\chi, \psi, \eta$ 
	and $\nu$ models. 
	The classification of the fields in the SO(10) and the SU(5) 
	groups is also shown. 
	The value of ${\rm U(1)'}$ charge follows the hypercharge 
	normalization. }
\begin{tabular}{cccccccc}  \hline \hline \\ [-0.4cm]
\vsp{0.2}
SO(10)& SU(5) & field & $Y$ & $2\sqrt{6}Q_\chi$ 
	& $\sqrt{72/5}Q_\psi$ & $Q_\eta$ 
	& $Q_\nu$ \\ \hline \\
\vsp{0.2}
{\bf 16} & {\bf 10} & 
$Q$     & $+\frac{1}{6}$ & $-1$& $+1$ & $-\frac{1}{3}$
	& $+\sqrt{\frac{1}{24}}$\\
& & 
$u^c$   & $-\frac{2}{3}$ & $-1$& $+1$& $-\frac{1}{3}$ 
	& $+\sqrt{\frac{1}{24}}$ \\
& & 
$e^c$   & $+1 $ & $-1$& $+1$& $-\frac{1}{3}$
	& $+\sqrt{\frac{1}{24}}$\\
\\
& $\bf{\ov{5}}$ &
$L$     & $-\frac{1}{2}$& +3 & $+1$& $+\frac{1}{6}$
	& $+\sqrt{\frac{1}{6}}$\\
& &
$d^c$   & $+\frac{1}{3}$ & $+3$& $+1$& $+\frac{1}{6}$
	& $+\sqrt{\frac{1}{6}}$ \\ 
\\
& {\bf 1} &
$\nu^c$ & $0$ & $-5$ & $+1$& $-\frac{5}{6}$& 0 \\ 
\\
{\bf 10} &  $\bf{5}$ &
$H_u$   & $+\frac{1}{2}$& +2 & $-2$& $+\frac{2}{3}$
	& $-\sqrt{\frac{1}{6}}$\\
& &
$D$     & $-\frac{1}{3}$ & $+2$& $-2$& $+\frac{2}{3}$
	& $-\sqrt{\frac{1}{6}}$ \\ 
\\ 
& $\bf{\ov{5}}$ &
$H_d$   & $-\frac{1}{2}$ & $-2$ & $-2$& $+\frac{1}{6}$
	& $-\sqrt{\frac{3}{8}}$\\
& &
$\ov{D}$& $+\frac{1}{3}$ & $-2$ & $-2$& $+\frac{1}{6}$
	& $-\sqrt{\frac{3}{8}}$ \\ 
\\
{\bf 1} & {\bf 1} & 
$S$     & $0$ & $0$ & $4$ & $-\frac{5}{6}$
	& $\sqrt{\frac{25}{24}}$ \\ 
\hline 
\hline 
\end{tabular} 
\label{table_u1charge}
\end{center}
\end{table}
Besides the SM quarks and leptons, there are two SM singlets 
$\nu^c$ and $S$, a pair of weak doublets $H_u$ and $H_d$, 
a pair of color triplets $D$ and $\ov{D}$ in each generation. 
The $\eta$ model arises when $E_6$ breaks into a rank-5 group 
directly in a specific compactification of the heterotic 
string theory~\cite{eta_ellis}. 
In the $\nu$ model, the right-handed neutrinos $\nu^c$ are 
gauge singlet~\cite{nu_model} and can have large Majorana 
masses to realize the see-saw mechanism~\cite{see-saw}. 

The ${\rm U(1)'}$ symmetry breaking occurs if the scalar 
component of the SM singlet field develops the vacuum 
expectation value (VEV). 
It can be achieved at near the weak scale via radiative 
corrections to the mass term of the SM singlet scalar 
field. 
Recent studies of the radiative ${\rm U(1)'}$ symmetry 
breaking can be found, \eg, in Ref.~\ref{ref:radiative_u1prime}. 

Several problems may arise in the $E_6$ models from view of 
low-energy phenomenology. For example, the presence of the baryon 
number violating operators give rise to too fast proton decay, 
or the absence of the Majorana neutrino mass terms (except for 
the $\nu$ model) requires a fine-tuning of the Dirac neutrino 
mass in order to satisfy experimentally observed neutrino mass 
relations. 
Some approaches to these problems are summarized in 
Ref.~\ref{ref:hewett_rizzo}.  
In the following, we assume that these problems are solved 
by an unknown mechanism. 
Moreover we assume that all the super-partners of the SM 
particles and the exotic matters do not affect the radiative 
corrections to the electroweak observables significantly,  
\ie, they are assumed to be heavy enough to decouple from 
the weak boson mass scale. 

\subsection{Phenomenological consequences of $Z$-$Z'$ mixing}
If the SM Higgs field carries a non-trivial ${\rm U(1)'}$ 
charge, its VEV induces the $Z$-$Z'$ mass mixing. 
On the other hand, the kinetic mixing between the hypercharge 
gauge boson $B$ and the ${\rm U(1)'}$ gauge boson $Z'$ 
can occur through the quantum effects below the GUT scale. 
After the electroweak symmetry is broken, the effective 
Lagrangian for the neutral gauge bosons in the 
${\rm SU(2)}_L \times {\rm U(1)}_Y \times {\rm U(1)'}$ 
theory is given by~\cite{eta_model} 
\bea
{\cal L}_{gauge} 
	&=&  -\frac{1}{4}Z^{\mu\nu}Z_{\mu\nu}
            -\frac{1}{4}Z'^{\mu\nu}Z'_{\mu\nu} 
	    -\frac{\sin \chi}{2}B^{\mu\nu}Z'_{\mu\nu}
	    -\frac{1}{4}A^{0\mu\nu}A^{0}_{\mu\nu} 
\nonumber \\ 
	& & + m^2_{ZZ'} Z^{\mu}Z'_{\mu}
	    +\frac{1}{2} m^2_Z Z^{\mu}Z_{\mu}
	    +\frac{1}{2} m^2_{Z'} Z'^\mu Z'_{\mu}, 
\label{eq:l_gauge}
\eea 
where $F^{\mu\nu} (F=Z,Z',A^0,B)$ represents the gauge field strength. 
The $Z$-$Z'$ mass mixing and the kinetic mixing are characterized 
by $m^2_{ZZ'}$ and $\sin \chi$, respectively. 
In this basis, the interaction Lagrangian for the neutral current 
process is given as 
\begin{eqnarray}
{\cal L}_{NC} &=& -\sum_{f,\, \alpha}  \left\{ 
	\; e \, Q_{f^{}_{\alpha}} \overline{f^{}_{\alpha}}
	\gamma^{\mu}f^{}_{\alpha} A^0_{\mu} +
	g^{}_Z \overline{f^{}_{\alpha}} \gamma^{\mu}
	\left( I^3_{f_L} - Q_{f^{}_{\alpha}} \sin^2\theta_W \right)
	f^{}_{\alpha} Z^{}_{\mu} \right. \nonumber \\ 
	& & \left. + g^{}_E Q^{f^{}_{\alpha}}_E 
	\overline{f^{}_{\alpha}}\gamma^{\mu}f^{}_{\alpha} 
	Z'_{\mu} \right\}, 
\label{eq:neutraC}
\end{eqnarray}
where $g_Z = g/\cos\theta_W = g_Y/\sin\theta_W$. 
The ${\rm U(1)'}$ gauge coupling constant is denoted by $g_E$ 
in the hypercharge normalization. 
The symbol $f_\alpha$ denotes the quarks or leptons with 
the chirality $\alpha$ ($\alpha = L$ or $R$). 
The third component of the weak isospin, the electric charge 
and the ${\rm U(1)'}$ charge of $f_\alpha$ are given by 
$I^3_{f_\alpha}$, $Q_{f_\alpha}$ and $Q_E^{f_\alpha}$, respectively. 
The ${\rm U(1)'}$ charge of the quarks and leptons listed in Table~1 
should be read as 
\bea
Q_E^Q = Q_E^{u_L} = Q_E^{d_L},~~Q_E^L = Q_E^{\nu_L} = Q_E^{e_L}, 
~~
Q_E^{f^c} = -Q_E^{f_R} ~~(f=e,u,d). 
\label{eq:charge_rule}
\eea
The mass eigenstates $(Z_1,Z_2,A)$ is obtained by 
the following transformation; 
\begin{equation}
\left( \begin{array}{c}  Z \\  Z' \\  A^0  
\end{array}\right) 
= 
\left(
	\begin{array}{ccc}
\cos \xi + \sin \xi \sin \theta_W \tan \chi &
-\sin \xi + \cos \xi \sin\theta_W \tan \chi & 0 \\
\sin \xi / \cos \chi & \cos \xi / \cos \chi & 0 \\
-\sin\xi \cos \theta_W \tan \chi & 
- \cos \xi \cos \theta_W \tan \chi & 1   
	\end{array}
\right) 
\left( \begin{array}{c} {Z_1} \\ 
{Z_2} \\ {A} \end{array}\right). 
\end{equation}
Here the mixing angle $\xi$ is given by 
\begin{equation}
\tan 2\xi = \frac{-2c^{}_{\chi}(m^2_{ZZ'}+s^{}_W s^{}_{\chi}
            m^2_Z)}{m^2_{Z'} - (c^2_{\chi}-s^2_W s^2_{\chi})m^2_Z+
            2s^{}_W s^{}_{\chi} m^2_{ZZ'}}~, 
\label{eq:angle_xi}
\end{equation}
with the short-hand notation, 
$c_\chi =  \cos\chi$, $s_\chi = \sin\chi$ and $s_W = \sin\theta_W$. 
The physical masses $m_{Z_1}$ and $m_{Z_2}$ ($m_{Z_1} < m_{Z_2}$) 
are given as follows; 
\bsub
\bea
m_{Z_{1}}^2 
	 &=& m_Z^2 (c_\xi + s_\xi s_W t_\chi)^2 
	+ m_{Z'}^2 \biggl( \frac{s_\xi}{c_\chi} \biggr)^2
	+ 2 m^2_{ZZ'} \frac{s_\xi}{c_\chi} (c_\xi + s_\xi s_W t_\chi),  
\label{eq:light_Z1}
\\
m_{Z_{2}}^2 
	 &=& m_Z^2 (c_\xi s_W t_\chi - s_\xi)^2 
	+ m_{Z'}^2 \biggl( \frac{c_\xi}{c_\chi} \biggr)^2
	+ 2 m^2_{ZZ'} \frac{c_\xi}{c_\chi} (c_\xi s_W t_\chi -s_\xi), 
\eea
\esub
where $c_\xi = \cos\xi$, $s_\xi = \sin\xi$ and $t_\chi = \tan\chi$. 
The lighter mass eigenstate $Z_1$ should be identified with 
the observed $Z$ boson at LEP1 or SLC. 
The excellent agreement between the current experimental results 
and the SM predictions at the quantum level implies that the 
mixing angle $\xi$ has to be small. 
In the limit of small $\xi$, the interaction Lagrangians 
for the processes 
$Z_{1,2} \to f_\alpha \ov{f_\alpha}$ are expressed as 
\begin{subequations}
\begin{eqnarray}
{\cal L}_{Z_1} &=& -\sum_{f,\, \alpha} g^{}_Z 
	\overline{f_{\alpha}} \gamma^{\mu} \left[
	\left( I^{3}_{f_{L}}-Q_{f_{\alpha}}\sin^2\theta_W 
	\right) + \tilde{Q}^{f_{\alpha}}_E \xibar \right]
	f_{\alpha}  Z_{1 \mu}, 
\label{eq:neutral1}\\ 
{\cal L}_{Z_2} &=& -\sum_{f,\,\alpha}\frac{g^{}_E}{c^{}_{\chi}}
	\overline{f_{\alpha}}\gamma^{\mu} \left[ \tilde{Q}^{f_{\alpha}}_E
	-\left( I^3_{f_{\alpha}} - Q_{f_{\alpha}}
	\sin^2\theta_W \right) \frac{g^{}_Z c^{}_{\chi}}{g^{}_E}
	\xi \right]f_{\alpha} Z_{2 \mu}, 
\label{eq:neutral2}
\end{eqnarray}
\label{eq:neutralboth}
\end{subequations}
\hsp{-0.3} 
where the effective mixing angle $\xibar$ 
in Eq.~(\ref{eq:neutral1}) is given as 
\beq
\xibar = \frac{g_E}{g_Z\cos \chi } \xi. 
\eeq
In Eq.~(\ref{eq:neutralboth}), the effective ${\rm U(1)'}$ 
charge $\tilde{Q}_E^{f_\alpha}$ is introduced as 
a combination of $Q_E^{f_\alpha}$ and the hypercharge $Y_{f_\alpha}$: 
\bsub
\bea
\tilde{Q}^{f_\alpha}_E &\equiv& Q^{f_\alpha}_E + Y_{f_\alpha} \delta, 
\label{eq:effective_charge}\\
\delta &\equiv&  -\frac{g^{}_Z}{g^{}_E}s^{}_W s^{}_{\chi}, 
\eea
\label{eq:u1_charge}
\esub
\hsp{-0.3}
where the hypercharge $Y_{f_\alpha}$ should be read from Table~1 
in the same manner with $Q_E^{f_\alpha}$ 
(see, Eq.~(\ref{eq:charge_rule})).  
As a notable example, one can see from Table~1 
that the effective charge $\tilde{Q}_E^{f_\alpha}$ of 
the leptons ($L$ and $e^c$) disappears in the 
$\eta$ model if $\delta$ is taken to be $1/3$~\cite{eta_model}. 

Now, due to the $Z$-$Z'$ mixing, the observed 
$Z$ boson mass $m_{Z_1}$ at LEP1 or SLC is shifted from 
the SM $Z$ boson mass $m_Z$: 
\beq
\Del m^2 \equiv m_{Z_1}^2 - m_Z^2 \le 0. 
\label{eq:mass_shift}
\eeq
The presence of the mass shift affects the 
$T$-parameter~\cite{peskin_takeuchi} at tree level. 
Following the notation of Ref.~\ref{ref:hhkm}, the $T$-parameter 
is expressed in terms of the effective form factors $\gzbar(0), 
\gwbar(0)$ and the fine structure constant $\alpha$:
\bsub
\bea
\alpha T &\equiv& 1 - \frac{\bar{g}^2_{W}(0)}{m^2_W}
	\frac{m^2_{Z_1}}{\bar{g}^2_Z(0)}  \\ 
	&=&
	\alpha \left(T_{\rm SM}^{}+  T_{\rm new}^{}\right), 
\eea
\esub
where 
$T_{\rm SM}^{}$ and the new physics contribution 
$T_{\rm new}$ are given by: 
\bsub
\bea
\alpha T_{\rm SM} 
	&=& 1 - \frac{\bar{g}^2_{W}(0)}{m^2_W}
		\frac{m^2_{Z}}{\bar{g}^2_Z(0)}, \\
\alpha T_{\rm new}
	& = & -\frac{ \Del m^2}{m^2_{Z_1}} \geq 0. 
\eea
\esub
It is worth noting that the sign of $T_{\rm new}$ is 
always positive. 
The effects of the $Z$-$Z'$ mixing in the $Z$-pole experiments 
have hence been parametrized by the effective mixing angle 
$\xibar$ and the positive parameter $T_{\rm new}$. 

We note here that we retain 
the kinetic mixing term $\delta$ as a part of the effective $Z_1$ 
coupling $\tilde{Q}_E^{f_\alpha}$ in Eq.~(\ref{eq:effective_charge}). 
As shown in Refs.~\ref{ref:eta_model},\ref{ref:general_zzmixing},
\ref{ref:holdom2}, 
the kinetic mixing term $\delta$ can be absorbed into a further 
redefinition of $S$ and $T$. 
Such re-parametrization may be useful 
if the term $Y_{f_\alpha} \delta$ in Eq.~(\ref{eq:effective_charge}) 
is much larger than the $Z'$ charge $Q_E^{f_\alpha}$. 
In the $E_6$ models studied in this paper, we find no merit in 
absorbing the $Y_f \delta$ term because, the remaining 
$Q_E^{f_\alpha}$ term is always significant. 
We therefore adopt $\tilde{Q}_E^{f_\alpha}$ as the effective 
$Z_1$ couplings and $T_{\rm new}$ accounts only for the 
mass shift (\ref{eq:mass_shift}). 
All physical consequences such as the bounds on $\xibar$ and 
$m_{Z_2}$ are of course independent of our choice of the 
parametrization. 

The two parameters $T_{\rm new}$ and $\xibar$ 
are complicated functions of the parameters of the effective 
Lagrangian (\ref{eq:l_gauge}). 
In the small mixing limit, 
we find the following useful expressions 
\bsub
\bea
\xibar &=& -\biggl( \frac{g_E}{g_Z}\frac{m_Z}{m_{Z'}} 
	\biggr)^2 \zeta \biggl[ 1+ O(\frac{m_Z^2}{m_{Z'}^2})\biggr], 
\\
\alpha T_{\rm new} &=& \hph \biggl( \frac{g_E}{g_Z}
		\frac{m_Z}{m_{Z'}} \biggr)^2 \zeta^2
	\biggl[ 1+ O(\frac{m_Z^2}{m_{Z'}^2})\biggr], 
\eea
\label{eq:tnew_xibar}
\esub
\hsp{-0.3}
where we introduced an effective mixing parameter $\zeta$ 
\beq
\zeta = \frac{g_Z}{g_E}\frac{m_{ZZ'}^2}{m_Z^2} - \delta.  
\label{eq:zeta}
\eeq
The $Z$-$Z'$ mixing effect disappears at $\zeta = 0$. 
Stringent limits on $m_{Z'}$ and hence on $m_{Z_2}$ 
can be obtained through the mixing effect if $\zeta$ is 
$O(1)$. 
We will show in Sec.~5 that $\zeta$ is calculable 
once the particle spectrum of the model is specified. 
The parameter $\zeta$ plays an essential role in 
the analysis of $Z'$ models. 

In the low-energy neutral current processes, effects of 
the exchange of the heavier mass eigenstate $Z_2$ can be 
detected. 
In the small $\xibar$ limit, they constrain the contact 
term $g_E^2/c_\chi^2 m_{Z_2}^2$. 

\section{Electroweak observables in the $Z'$ model}
	\begin{table}[]
\tcaption{\small 
Electroweak measurements at LEP, SLC, Tevatron and LENC experiments. 
The average $W$-boson mass is found in Ref.~\ref{ref:wboson_moriond}.  
Except for the weak charge of cesium atom~\cite{apv_data}, the data of 
LENC experiments given in this table, which have been reduced from 
the original data~\cite{slac,cern,bates,mainz,fh,ccfr,charm-II},  
is summarized in Refs.~\ref{ref:chu},\ref{ref:chm}. 
The pull factors are given at the best fit point of 
the SM and four $Z'$ models at $\mh=100\gev$ and a constraint 
$\tnew \ge 0$. 
The best fit values of parameters in both the SM and $Z'$ models 
are shown in Table~\ref{tab:best_fit}. 
Correlation matrix elements of the $Z$ line-shape parameters 
and those for the heavy-quark parameters 
are found in Ref.~\ref{ref:lepewwg98}. 
The data $R_\ell^{}$ and $A^{0,\ell}_\fb$ are obtained 
by assuming the $e$-$\mu$-$\tau$ universality. 
}
	\begin{center}
	\begin{tabular}{|r|c|c|c|c|c|c|}
	\hline
	 & data  & SM & $\chi$ & $\psi$& $\eta$ & $\nu$ 
	\\ \hline
	\makebox[35mm][l]{{\bf $Z$ parameters~\cite{lepewwg98}
	}} &&&&&& \\
	$m^{}_Z$ (GeV) & 91.1867 $\pm$  0.0021 &------ 
	&------ &------ &------ &------ \\
	$\Gamma_Z^{}$ (GeV)  & $2.4939\pm 0.0024$  & $-1.0$ 
	& $-1.1$ & $-1.0$ & $-0.9$ & $-1.0$ \\
	$\sigma^0_h$(nb) & $41.491\pm 0.058\hpz$ & $\hph 0.4$ 
	& $\hph 0.4$ & $\hph 0.6$ & $\hph 0.4$& $\hph 0.5$\\
	$R_{\ell}$ & $20.765\pm 0.026\hpz$ & $\hph 0.5$ 
	& $\hph 0.5$ & $\hph 0.3$ & $\hph 0.5$& $\hph 0.5$\\
	$A^{0,\ell}_\fb$ & $0.01683\pm 0.00096$ & $\hph 0.6$ & 
	$\hph 0.6$ & $\hph 0.6$& $\hph 0.6$& $\hph 0.6$\\
	$A_{\tau}$& 0.1431 $\pm$ 0.0045 & $-0.9$ & $-0.9$
	& $-0.9$& $-0.9$ & $-0.9$\\
	$A_{e} $  & 0.1479 $\pm$ 0.0051 & $\hph0.1$ & $\hph 0.2$
	& $\hph 0.1$& $\hph 0.1$& $\hph 0.2$\\
	$R_b$ & 0.21656 $\pm$ 0.00074 & $\hph 1.0$ & $\hph 1.0$
	& $\hph 1.1$& $\hph 1.1$& $\hph 1.1$ \\
	$R_c$ & 0.1735 $\pm$ 0.0044 & $\hph 0.3$ & $\hph 0.3$
	& $\hph 0.3$& $\hph 0.3$& $\hph 0.3$\\
	$A^{0,b}_{FB}$ & 0.0990 $\pm$ 0.0021 & $-2.0$ & $-1.9$
	& $-2.1$& $-2.0$& $-2.0$\\
	$A^{0,c}_{FB}$ & 0.0709 $\pm$ 0.0044 & $-0.7$ & $-0.6$
	& $-0.7$& $-0.7$ & $-0.6$\\
	$A^0_{LR}$ & 0.1510 $\pm$ 0.0025 & $\hph 1.5$ & $\hph 1.6$
	& $\hph 1.5$ & $\hph 1.5$& $\hph 1.6$\\
	$A_b$ & $0.867\pm 0.035$ & $-1.9$ & $-1.9$& $-1.9$
	& $-1.9$& $-1.9$ \\
	$A_c$ & $0.647\pm 0.040$ & $-0.5$ & $-0.5$& $-0.5$
	& $-0.5$& $-0.5$\\
	\hline	
	\makebox[35mm][l]{{\bf $W$ mass~\cite{wboson_moriond}
	}}&&&&&& \\
	$m^{}_W$ (GeV) & 80.410 $\pm$ 0.044 & $\hph 0.8$ & $\hph 0.8$ 
	& $\hph 0.8$& $\hph 0.8$ & $\hph 0.8$\\
	\hline	
	\makebox[35mm][l]{{\bf LENC exp.~\cite{chm,chu} }}&&&&&&\\
	$A_{\rm SLAC}$ & $\hph$0.80 $\pm$ 0.058& $\hph 1.0$ 
	& $\hph 1.0$ & $\hph 1.0$& $\hph 1.1$
	& $\hph 1.0$	\\
	$A_{\rm CERN}$ &$-$1.57 $\pm$ 0.38 & $-0.4$ & $-0.4$
	& $-0.4$& $-0.4$& $-0.4$ \\
	$A_{\rm Bates}$ & $-$0.137 $\pm$ 0.033 & $\hph 0.5$ 
	& $\hph0.4$ & $\hph0.5$& $\hph0.4$& $\hph0.4$\\
	$A_{\rm Mainz}$ & $-$0.94 $\pm$ 0.19\,\,\,& $-0.3$ & $-0.3$
	& $-0.3$ & $-0.4$& $-0.3$\\
	$Q_W(^{133}_{55}Cs)$~\cite{apv_data} 
	  & $-$72.06 $\pm$ 0.44 \hpz\,\,  
	& $\hpz	2.2$ & $-0.1$ & $\hph 2.2$& $\hph 0.1$
	& $\hph 0.0$ \\
	$K_{FH}$ & 0.3247 $\pm$ 0.0040  & $-1.5$ & $-1.4$ 
	& $-1.5$ & $-1.5$ & $-1.4$ \\
	$K_{CCFR}$ & 0.5820 $\pm$ 0.0049& $-0.5$ & $-0.4$& $-0.4$
	& $-0.5$& $-0.4$\\
	$g_{LL}^{\nu_\mu e}$  & $-$0.269 $\pm$ 0.011 \hpz & $\hph
	0.4$ & $\hph0.1$ & $\hph0.1$ & $\hph0.4$& $\hph0.1$\\
	$g_{LR}^{\nu_\mu e}$  & 0.234 $\pm$ 0.011 & $\hph 0.1$ 
	& $\hph0.0$ & $\hph0.4$& $\hph0.2$& $\hph0.2$\\
	\hline	
	$\chi^2_{\rm min}$ & & 23.8& 18.3 & 23.5& 19.2& 18.4\\
	\hline	
	\end{tabular} 
\label{tab:ewdata}
	\end{center}
\end{table}
In this section, we briefly discuss the theoretical 
framework~\cite{chu,uch} to calculate the electroweak observables 
which are used in our analysis. 
The experimental data of the $Z$-pole experiments, the $W$-boson 
mass measurement and the low-energy experiments 
used in this paper are summarized in Table~\ref{tab:ewdata}. 

The pseudo-observables of the $Z$-pole experiments are expressed 
in terms of the effective coupling $g_\alpha^f$~\cite{lepewwg98}, 
where $f$ denotes all the SM fermions except for the top-quark, 
and $\alpha$ being their chirality, $L$ or $R$. 
Following our parametrization of the $Z$-$Z'$ mixing 
(\ref{eq:neutral1}), the effective coupling $g^f_{\alpha}$ 
in the $Z'$ models can be expressed as 
\bea
g_\alpha^f = (g_\alpha^f)_{\rm SM} + \tilde{Q}_E^{f_\alpha} \xibar. 
\eea
The SM prediction for the effective coupling $(g_\alpha^f)_{\rm SM}$ 
can be expanded in terms of the gauge boson propagator corrections 
$\dgzbar$ and $\dsbar$:
\bea
(g_\alpha^f)_{\rm SM} = a + b \dgzbar + c\dsbar, 
\label{eq:eff_coupling}
\eea
where the numerical coefficients $a,b$ and $c$ are given in 
Refs.~\ref{ref:chu},\ref{ref:uch}. 
Two parameters $\dgzbar$ and $\dsbar$ in Eq.~(\ref{eq:eff_coupling}) 
are defined as the shift in the effective couplings 
$\gzbar(m_{Z_1}^2)$ and $\sbar(m_{Z_1}^2)$~\cite{hhkm} 
from their SM reference values at $\mt = 175~{\rm GeV}$ and 
$\mh = 100~{\rm GeV}$. 
They can be expressed in terms of the $S$ and $T$ 
parameters~\cite{peskin_takeuchi} as 
\bsub
\begin{eqnarray}
\!\!\!\!\!\!\!\! \Del \bar{g}^2_Z &=&  \gzbar(m_{Z_1}^2) - 0.55635 
	= 0.00412 \dt + 0.00005[1-(100~{\rm GeV}/m^{}_H)^2], \\  
\!\!\!\!\!\!\!\! \Del \bar{s}^2 &=& \sbar(m_{Z_1}^2) - 0.23035 
	= 0.00360 \ds - 0.00241 \dt - 0.00023 \xa, 
\end{eqnarray}
\esub
where the expansion parameter $\xa$ is introduced to estimate 
the uncertainty of the hadronic contribution to the QED coupling 
$1/\ov{\alpha}(m_{Z_1}^2) = 128.75 \pm 0.09$~\cite{EJ}:
\beq
\xa \equiv \frac{1/\ov{\alpha}(m_{Z_1}^2) - 128.75}{0.09}. 
\label{eq:xa_qed}
\eeq
Here, $\ds,\dt,\du$ parameters are also measured from their 
SM reference values and they are given as the sum of the SM 
and the new physics contributions 
\begin{equation}
\ds = \ds_{\rm SM}+S^{}_{\rm new}, \;\;
\dt = \dt_{\rm SM}+T^{}_{\rm new}, \;\; 
\du = \du_{\rm SM}+U^{}_{\rm new}. 
\label{eq:stu_delta}
\end{equation}
A convenient parametrization of $\ds_{\rm SM},\dt_{\rm SM}$ and 
$\du_{\rm SM}$ in terms of $\mt$ and $\mh$ has been given 
in Ref.~\ref{ref:hhm}. 
The formulae of the $Z$-pole observables listed in 
Table~\ref{tab:ewdata} in terms of $g_\alpha^f$ can be found in 
Refs.~\ref{ref:chu},\ref{ref:uch}. 

The theoretical prediction of $\mw$ is given as~\cite{hhkm,hhm}
\begin{equation}
\mw(\gev) =80.402-0.288\,\ds +0.418\,\dt +0.337\,\du
	+0.012\,\xa, 
\end{equation}
by using the same parameters, $\ds, \dt, \du$ (\ref{eq:stu_delta}) 
and $\xa$ (\ref{eq:xa_qed}). 

The observables in the LENC experiments which are used 
in our analysis are as follows -- 
(i) polarization asymmetry of the charged lepton scattering off 
nucleus target, 
(ii) parity violation in cesium atom, 
(iii) inelastic $\nu_\mu$-scattering off nucleus target and 
(iv) neutrino-electron scattering. 
Theoretical expressions for the observables of 
(i) and (ii) are conveniently given in terms of 
the model-independent parameters $C_{1q}, C_{2q}$~\cite{jekim}  
and $C_{3q}$~\cite{chm}.  
The $\nu_\mu$-scattering data (iii) and (iv) are expressed 
in terms of the parameters $g_{L\alpha}^{\nu_\mu f}$. 
In the $Z'$ models, these model-independent parameters can be 
written as follows: 
\bsub
\bea
C_{iq} &=& (C_{iq})_{\rm SM} + \Del C_{iq}, \\
g_{L\alpha}^{\nu_\mu f} &=& (g_{L\alpha}^{\nu_\mu f})_{\rm SM} + 
	\Del g_{L\alpha}^{\nu_\mu f}, 
\eea
\esub
where the first term in each equation is the SM contribution
which is parametrized conveniently by $\ds$ and $\dt$~\cite{chm}. 
The second terms $\Del C_{iq}$ and $\Del g_{L\alpha}^{\nu_\mu f}$ 
represent the additional contributions from the $Z$-$Z'$ mixing 
and the direct $Z_2$ exchange, which are proportional to 
$\xibar$ and $\contact$, respectively. 
The theoretical prediction of the LENC observables in terms 
of $\Del C_{iq}$ and $\Del g_{L\alpha}^{\nu\mu f}$ 
can be found in Ref.~\ref{ref:chu}. 

\section{Constraints on $Z'$ bosons from electroweak experiments}

\subsection{$\chi^2$-analysis on the $Z'$ models}
There are six free parameters in the $Z'$ models -- 
the tree level contribution to the $T$ parameter $\tnew$, 
the $Z$-$Z'$ mass mixing angle $\xibar$, the direct $Z_2$-boson 
contribution to the low-energy processes $\contact$, 
and the three SM parameters, 
$\mt, \alpha_s(m_{Z_1})$ and $\abar(m_{Z_1}^2)$. 
Throughout our analysis, we use 
$\mt = 173.8 \pm 5.2~({\rm GeV})$~\cite{PDG98}, 
$\alpha_s (m_{Z_1}^{}) = 0.119 \pm 0.002$~\cite{PDG98}, 
and 
$1/\abar(m_{Z_1}^2) = 128.75 \pm 0.09$~\cite{EJ} 
as constraints on the SM parameters. 
The Higgs mass dependence of the results are parametrized 
by $\xh \equiv \ln(\mh/100\gev)$ 
in the range $90~{\rm GeV} <  \mh \lsim 150~{\rm GeV}$. 
The lower bound is obtained at the LEP2 
experiment~\cite{higgs_moriond}. 
The upper bound is the theoretical limit on the lightest Higgs 
boson mass in any supersymmetric models that accommodate 
perturbative unification of the gauge couplings~\cite{kane}. 
	\begin{table}[t]
\tcaption{\small 
Summary of the best fit values of the parameters in the SM and 
four $Z'$ models $(\delta=0)$ for $\mh=100\gev$ and $\tnew\ge 0$. 
The mixing parameter $\xibar$ and the contact term $\contact$ 
are given in units of $10^{-4}$ and $\tev^{-2}$, respectively. 
The constraints on $\mt$ and $\alpha_s(m_{Z_1})$ are found in  
Ref.~\ref{ref:PDG98} while that on $\abar(m_{Z_1}^2)$ is 
given in Ref.~\ref{ref:EJ}. 
}
	\begin{center}
	\begin{tabular}{|r|c|c|c|c|c|c|}
	\hline
	Parameters & Constraints & SM & $\chi$ & $\psi$ 
	& $\eta$ & $\nu$ \\ \hline
	$\mt$ (GeV) & 173.8 $\pm$ 5.2\hpz\hpz 
	& 170.5 & 171.3 & 170.8 & 170.9 & 171.3 \\
	$\alps(m_{Z_1})$ & 0.119 $\pm$ 0.002 & 0.1186 
	& 0.1185 & 0.1184 & 0.1186 & 0.1185 \\
	$1/\abar(m_{Z_1}^2)$ & 128.75 $\pm$ 0.09 \hpz& 128.76 
	& 128.74 & 128.75 & 128.73 & 128.74 \\
	$\tnew$ & --- & --- & 0 & 0 & 0 & 0 \\
	$\xibar(10^{-4})$ & --- & --- & $0.32$ & $1.63$ & $-2.66$
	& $0.68$ \\
	$\contact $ & --- & --- &0.245&1.41&$-0.839$ & $0.635$\\
	\hline
	\end{tabular} 
\label{tab:best_fit}
	\end{center}
	\end{table}

We summarize the results of the fit for the $\psi, \chi, \eta$ 
and $\nu$ models: 
\bsub
\bea
&&
(1)~\chi~{\rm model}~(\delta=0)\nonumber \\
\vsk{0.1}
&&
\left.
\begin{array}{ll}
\tnew &=  -0.059 + 0.14 \xh \pm 0.098 \\
\xibar(10^{-4}) &= 0.02 - 9\xh \pm 4.05  \\
\contact &= 0.237 + 0.0032 \xh \pm 0.107 
\end{array}
\right \}, 
\chi^2_{\rm min} = 17.8 + 1.2 \xh,~ \\
\vsk{0.2}
&&
(2)~\psi~{\rm model}~(\delta=0)\nonumber \\
\vsk{0.1}
&&
\left.
\begin{array}{ll}
\tnew &=  -0.075 + 0.14 \xh \pm 0.097 \\
\xibar(10^{-4}) &= 1.2 - 1.3 \xh \pm 4.8  \\
\contact &= 1.31 + 0.16 \xh \pm 2.97
\end{array}
\right \}, 
\chi^2_{\rm min} = 22.9 + 1.4 \xh,~ \\
\vsk{0.2}
&&
(3)~\eta~{\rm model}~(\delta=0)\nonumber \\
\vsk{0.1}
&&
\left.
\begin{array}{ll}
\tnew &=  -0.062 + 0.14 \xh \pm 0.097 \\
\xibar(10^{-4}) &= -2.7 - 6.7 \xh \pm 9.4  \\
\contact &= -0.814 + 0.089 \xh \pm 0.449
\end{array}
\right \}, 
\chi^2_{\rm min} = 18.7 + 0.7 \xh,~ \\
\vsk{0.2}
&&
(4)~\nu~{\rm model}~(\delta=0)\nonumber \\
\vsk{0.1}
&&
\left.
\begin{array}{ll}
\tnew &=  -0.057 + 0.14 \xh \pm 0.098 \\
\xibar(10^{-4}) &= -0.42 + 0.6 \xh \pm 3.9 \\
\contact &= 0.619 + 0.024 \xh \pm 0.275
\end{array}
\right \}, 
\chi^2_{\rm min} = 18.0 + 1.1 \xh,~ 
\eea
\esub
where $\dof = 20$. 
	\begin{figure}[t]
	\begin{center}
	\leavevmode\psfig{figure=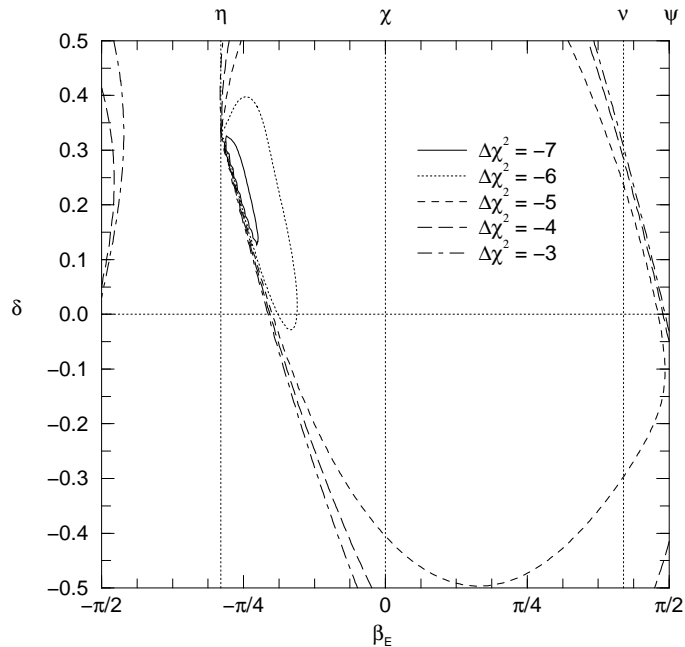,width=9cm}
	\end{center}
	\caption{Contour plot of 
	$\Del \chi^2 \equiv \chi_{\rm min}^2(\beta_E,\delta) 
	- \chi^2_{\rm min}({\rm SM})$ for $\mh = 100~{\rm GeV}$. 
	The mixing angle $\beta_E$ for the $\chi,\psi,\eta$ and 
	$\nu$ models are shown by vertical dotted lines. 
	The step of each contour is 1. }
	\label{chisq_distribution}
	\end{figure}
The mixing angle $\xibar$ and the contact term $\contact$ 
are given in units of $10^{-4}$ and $\tev^{-2}$, respectively. 
The best fit value of $\tnew$ falls into the unphysical region 
($\tnew < 0$) for all $Z'$ models even if the Higgs boson mass 
is its upper limit ($\sim 150\gev$). 
It should be noticed that $\tnew$ and $\xibar$ are consistent with 
zero in all models while $\contact$ shows the deviation from zero 
in the 1-$\sigma$ level for $\chi,\eta,\nu$ models. 

The best fit values of the six-parameters for $\mh = 100~{\rm GeV}$ 
under the condition $T_{\rm new} \geq 0$ are shown in 
Table~\ref{tab:best_fit}, together with the SM best fit 
result at $\mh=100\gev$. 
Only the best fit value of $\contact$ in the $\eta$ model is found 
in the unphysical region ($\contact<0$). 
The pull factors of the electroweak observables at the best fit 
point are also shown in Table~\ref{tab:ewdata}. 
We learn from the table that almost no improvement of the fit over 
the SM is found for the $Z$-pole and $\mw$ measurements. 
However, the $\chi,\eta, \nu$ models show the excellent fit 
to the weak charge of cesium atom $Q_W(^{133}_{55}Cs)$: 
the pull factor is reduced from 2.2 (SM) to less than 0.1. 
This may imply that more than 2-$\sigma$ deviation of 
the APV data from the SM prediction could be explained by the 
direct exchange 
of the $Z_2$ boson in the low-energy processes~\cite{apv_zprime}. 
On the other hand, the $\psi$ model does not show the reduction 
of the pull factor in $Q_W(^{133}_{55}Cs)$. 
Since all the SM matter fields in the $\psi$ model have the same 
${\rm U(1)'}$ charge (see Table~\ref{table_u1charge}), 
the couplings of contact interactions are parity conserving, 
which makes the contact term useless in the fit to the APV. 

We introduce a parameter 
\bea
\Del \chi^2 \equiv \chi^2_{\rm min}(\beta_E,\delta) 
	- \chi^2_{\rm min}(\rm SM),  
\eea
to measure the goodness of the fit in the $Z'$ models compared to 
the SM. 
We can see from Table~\ref{tab:ewdata} that the $\chi,\eta$ and 
$\nu$ models lead to $\Del \chi^2 = -5.5(\chi$), $-4.6(\eta)$ 
and $-5.4(\nu)$, respectively while the $\psi$ model shows no 
improvement of the fit, $\Del \chi^2 = -0.3$.  

	\begin{figure}[t]
	\begin{center}
	\leavevmode\psfig{figure=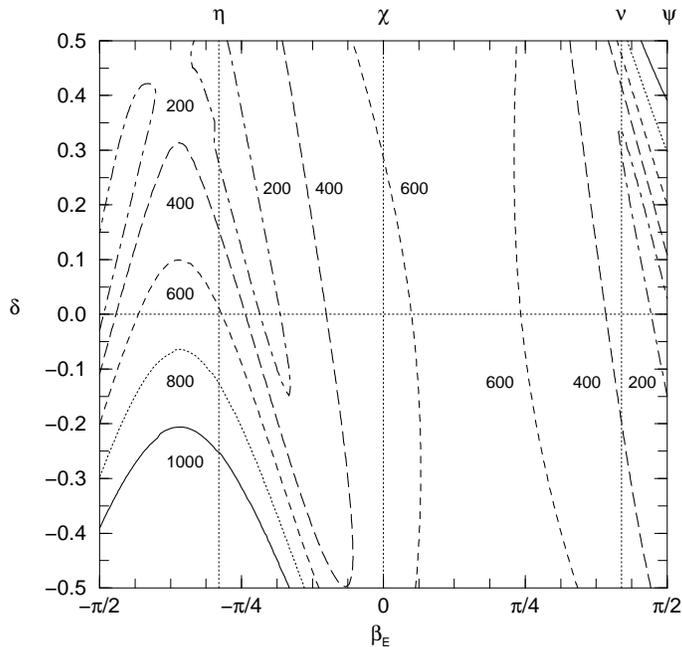,width=9cm}
	\end{center}
	\caption{Contour plot of the 95\% CL lower mass 
	limit of the $Z_2$ boson obtained from the LENC experiments 
	for $g_E = g_Y$ and $\mh = 100~{\rm GeV}$. 
	The vertical dotted lines correspond to 
	the $\chi, \psi, \eta$ and $\nu$ models.
	The limits are given in unit of GeV.}
	\label{mass_distribution}
	\end{figure}
In order to see the impact of kinetic mixing on the fit, 
we show the contour plot of $\Del \chi^2$ from the 
electroweak data under the conditions $\tnew,\contact \ge 0$ 
on the $(\beta_E,\delta)$ plane in Fig.~\ref{chisq_distribution}. 
We can see from the figure that the fit of 
the $\eta$ model at $\delta=0$ is rather worsen 
($\Del \chi^2 \sim -3$) as compared to that given in 
Table~\ref{tab:ewdata} ($\Del \chi^2 = -4.6$) because the fit 
in the table has been 
found without imposing the constraint $\contact \ge 0$.  
The leptophobic $\eta$ model ($\delta = 1/3$) does not improve 
the fit because, due to the leptophobity, the model does not have 
the contact term $\contact$ which is used to make the fit to 
the LENC (essentially the APV) data better. 
We find that the $Z'$ model with $(\beta_E, \delta) \approx 
(-\pi/4,0.2)$ shows the most excellent fit over the SM where 
$\Del\chi^2 \lsim -7$. 

\subsection{Lower mass bound on $Z'$ bosons}
As we expected from the formulae for $T_{\rm new}$ and $\xibar$ 
in the small mixing limit (\ref{eq:tnew_xibar}), 
the $Z_2$ mass is unbounded from the $Z$-pole data at $\zeta = 0$. 
For models with very small $\zeta$, the lower bound of the heavier 
mass eigenstate $Z_2$ in the $Z'$ models, therefore, comes from 
the LENC experiments.  
In Fig.~\ref{mass_distribution}, we show the contour plot of 
the 95\% CL lower mass limit of $Z_2$ boson from the LENC 
experiments on the $(\beta_E, \delta)$ plane by setting 
$g_E = g_Y$ and $\mh = 100~{\rm GeV}$ under the condition 
$m_{Z_2} \geq 0$. 
In practice, we obtain the 95\% CL lower limit of the $Z_2$ boson 
mass $m_{95}$ in the following way: 
\bea
0.05 =
\frac{
\disp{
\int^\infty_{g_E^2/m_{95}^2} 
	d\biggl(\frac{g_E^2}{m_{Z_2}^2}\biggr)
	P\biggl(\frac{g_E^2}{m_{Z_2}^2}\biggr)
}}
{\disp{
\int^\infty_{0} 
	d\biggl(\frac{g_E^2}{m_{Z_2}^2}\biggr)
	P\biggl(\frac{g_E^2}{m_{Z_2}^2} \biggr)}
}, 
\eea
where we assume that the probability density function 
$P(g_E^2/m_{Z_2}^2)$ is proportional to 
${\rm exp}(-\chi^2(g_E^2/m_{Z_2}^2)/2)$. 

We can read off from Fig.~\ref{mass_distribution} that 
the lower mass bound of the $Z_2$ boson in the $\psi$ model 
at $\delta = 0$ is much weaker than those of the other $Z'$ 
models. 
This is because, as we mentioned before, the U(1)$'$ charge 
assignment on the SM matter fields in the model makes the 
constraint from the APV measurement useless. 
We also find in Fig.~\ref{mass_distribution} that the lower mass 
bound of the $Z_2$ boson disappears near the leptophobic $\eta$-model 
($\beta_E = \tan^{-1}(\sqrt{5/3})$ and 
$\delta = 1/3$)~\cite{eta_model}. 
Furthermore the lower mass bound tend to be small at the 
``best fit'' point which we found in Fig.~\ref{chisq_distribution},  
$(\beta_E,\delta)=(-\pi/4, 0.2)$. 

We summarize the 95\% CL lower bound on $m_{Z_2}$ 
for the $\chi,\psi,\eta$ and $\nu$ models ($\delta = 0$) 
in Table~\ref{mzelenc}. 
For comparison, those in Ref.~\ref{ref:chu} are given in the 
same table. 
It should be noticed that the bounds on $Z_\chi,Z_\eta$ and $Z_\nu$ 
masses are more severely constrained as compared to 
Ref.~\ref{ref:chu} 
due to the improved value of $Q_W(^{133}_{55}Cs)$ while the bound 
on the $Z_\psi$ mass is almost unchanged. 
\begin{table}[t]
\begin{center}
\tcaption{The 95\% CL lower bound of $m_{Z_2}$ (GeV) in the 
	$\chi, \psi, \eta$ and $\nu$ models ($\delta = 0$) 
	for $g_E=g_Y$ and $\mh = 100~{\rm GeV}$. 
	Those in Ref.~\ref{ref:chu} and 
	the result of direct search~\cite{direct_search} are 
	shown for comparison. }
\begin{tabular}{c|cccc} \hline \hline 
	& $\chi$ & $\psi$ & $\eta$ & $\nu$ \\ \hline
our results & 554 & 137 & 619 & 342 \\ 
results in Ref.~\ref{ref:chu}& 451 & 136 & 317 & 284 \\ 
direct search~\cite{direct_search} 
	& 595 & 590 & 620 & --- 
\\ \hline \hline 
\end{tabular}
\label{mzelenc}
\end{center}
\end{table}  

\begin{figure}[t]
\begin{center}
	\leavevmode\psfig{figure=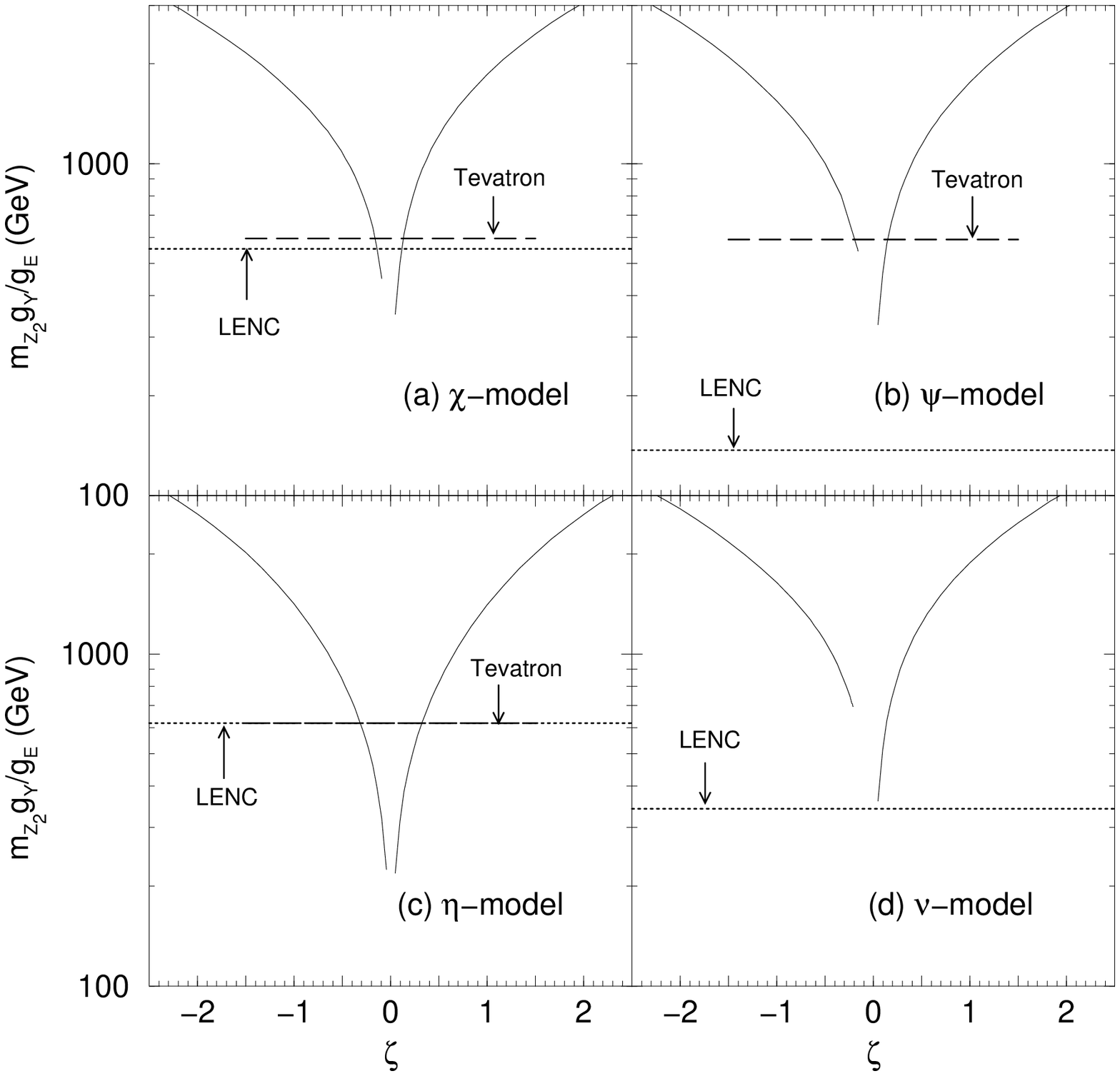,width=9cm}
\fcaption{The 95\% CL lower mass limit of $Z_2$ in the 
	$\chi$, $\psi$, $\eta$ and $\nu$ models 
	for $\mh = 100~{\rm GeV}$. 
	The $Z_2$ boson mass is normalized by $g_E/g_Y$. 
	Constraints from $Z$-pole experiments and LENC 
	experiments are separately shown. 
	The results of the direct search at Tevatron~\cite{direct_search}
	for the $\chi,\psi$ and $\eta$ models are also 
	shown.  
	}
\label{mass_95cl}
\end{center}
\end{figure}

We have found that the $Z$-pole, $\mw$ and the LENC data 
constrain ($T_{\rm new}, \xibar$), $T_{\rm new}$ and $\contact$, 
respectively. 
We can see from Eq.~(\ref{eq:tnew_xibar}) that, for a given $\zeta$, 
constraints on $T_{\rm new}, \xibar$ and $\contact$ can be 
interpreted as the bound on $m_{Z_2}$. 
We show the 95\% CL lower mass bound of the $Z_2$ boson 
for $\mh = 100~{\rm GeV}$ in four $Z'$ models as a function of $\zeta$. 
The bound is again found under the condition $m_{Z_2} \geq 0$. 
Results are shown in Fig.~\ref{mass_95cl}(a) $\sim$  
\ref{mass_95cl}(d) for the $\chi,\psi,\eta,\nu$ models, 
respectively. 
The lower bound from the $Z$-pole and $\mw$ data, and 
that from the LENC data are separately plotted in the same figure. 
Shown in the figure is the lower bound of $m_{Z_2}g_Y/g_E$ 
for $g_E = g_Y$. 
The bound on $m_{Z_2}g_Y/g_E$ is approximately independent of 
$g_E$ for $g_E/g_Y = 0.5 \sim 2.0$ in each model~\cite{chu}. 
The $Z_2$ mass is unbounded from the $Z$-pole data at $\zeta = 0$ 
because the data constrain $\tnew$ and $\xibar$ which are proportional 
to $\zeta^2$ and $\zeta$, respectively. 
Then, the lower bound on $m_{Z_2}$ at very small $\zeta$ is obtained 
from the LENC experiments and the direct search experiment at 
Tevatron. 
For comparison, we plot the 95\% CL lower bound on 
$m_{Z_2}$ obtained from the direct search experiment~\cite{direct_search} 
in Fig.~\ref{mass_95cl}. 
In the direct search experiment, the $Z'$ decays into the exotic 
particles, \eg, the decays into the light right-handed neutrinos 
which are expected for some models, are not taken into account. 
We summarize the 95\% CL lower bound on $m_{Z_2}$ 
for the $\chi,\psi,\eta$ and $\nu$ models ($\delta = 0$) 
obtained from the low-energy data and the direct search 
experiment~\cite{direct_search} in Table~\ref{mzelenc}. 
The lower bound of $m_{Z_2}$ in the $\eta$ model from the LENC 
experiments is competitive the bound from the direct search 
experiment. 

The lower bound of $m_{Z_2}$ is affected by the Higgs boson mass 
through the $T$ parameter. 
As we mentioned previously, 
$T_{\rm new}$ tends to be in the physical region ($T_{\rm new} \geq 0$) 
for large $\mh$ $(\xh)$. 
Then, we find that the large Higgs boson mass decreases the lower 
bound of $m_{Z_2}$. 
For $\zeta = 1$, the lower $m_{Z_2}$ bound in the $\chi,\psi,\nu$ 
($\eta$) models for $\mh = 150~{\rm GeV}$ is weaker than that for 
$\mh = 100~{\rm GeV}$ about 7\% (11\%). 
On the other hand, the Higgs boson with $\mh = 80~{\rm GeV}$ 
makes the lower $m_{Z_2}$ bound in all the $Z'$ models 
severe about 5\% as compared to the case for $\mh = 100~{\rm GeV}$.  
Because $T_{\rm new}$ and $\xibar$ are proportional to 
$\zeta^2$ and $\zeta$, respectively (see Eq.~(\ref{eq:tnew_xibar})), 
and it is unbounded at $|\zeta| \simeq 0$, 
the lower bound of $m_{Z_2}$ may be independent of $\mh$ 
in the small $|\zeta|$ region. 
The $\mh$-dependence of the lower mass bound obtained 
from the LENC data is safely negligible. 

It has been discussed that the presence of $Z_2$ boson 
whose mass is much heavier than the SM $Z$ boson mass,  
say 1 TeV, may lead to a find-tuning problem to stabilize 
the electroweak scale against the ${\rm U(1)'}$ scale~\cite{drees}. 
The $Z_2$ boson with $m_{Z_2} \leq 1~{\rm TeV}$ for $g_E = g_Y$ 
is allowed by the electroweak data only if $\zeta$ satisfies 
the following condition: 
\bea 
\begin{array}{ll}
-0.5 \lsim \zeta \lsim +0.4 & ~~{\rm for~the}~\chi,\psi,\nu~{\rm models}, 
\\
-0.6 \lsim \zeta \lsim +0.6 & ~~{\rm for~the}~\eta~{\rm model}. 
\end{array}
\label{eq:zeta_condition}
\eea

\section{Light $Z'$ boson in minimal SUSY $E_6$-models}
It is known that the gauge couplings are not unified in the $E_6$ 
models with three generations of {\bf 27}. 
In order to guarantee the gauge coupling unification, a pair of 
weak-doublets, $H'$ and $\ov{H'}$, should be added into  
the particle spectrum at the electroweak scale~\cite{dienes}. 
They could be taken from ${\bf 27} + {\bf \ov{27}}$ or the adjoint 
representation {\bf 78}. 
The ${\rm U(1)'}$ charges of the additional weak doublets should 
have the same magnitude and opposite sign, $a$ and $-a$,  to cancel 
the ${\rm U(1)'}$ anomaly. 
In addition, a pair of the complete SU(5) multiplet such as 
${\bf 5 + \ov{5}}$ can be added without spoiling the unification 
of the gauge couplings~\cite{eta_model,dienes}. 
\begin{table}[t]
\begin{center}
\tcaption{Coefficients of the 1-loop $\beta$-functions 
	for the gauge couplings in the MSSM and the minimal 
	$E_6$ models. 
	The model $\chi(16)$ has three generations of ${\bf 16}$ 
	and a pair ${\bf 2}+{\bf\ov{2}}$. 
	The model $\chi(27)$ and $\psi, \eta,\nu$ have three 
	generations of ${\bf 27}$ and a pair ${\bf 2}+{\bf\ov{2}}$. 
	}
\begin{tabular}{ccccccc}
\\ \hline \hline 
	& MSSM & $\chi(16)$ & $\chi(27)$ & $\psi$ & $\eta$ & $\nu$ 
\\ \hline 
$b_1$	& $\frac{33}{5}$ & $\hph \frac{33}{5}$ 
	& $\frac{48}{5}$ & $\frac{48}{5}$ 
	& $\frac{48}{5}$ & $\frac{48}{5}$ \\
$b_2$	& 1    & \hph 1 & 4 & 4 & 4 & 4 \\
$b_3$	& $-3$ & $-3$ & 0 & 0 & 0 & 0 \\
$b_E$	& ---  & $6+ \frac{a^2}{10}$ & $9+ \frac{a^2}{10}$ 
	& $9+ \frac{a^2}{6}$ & $9+ \frac{12}{5}a^2$ 
	& $9+ \frac{12}{5}a^2$ \\
$b_{1E}$& ---  & $-\sqrt{\frac{3}{50}}a$ & $-\sqrt{\frac{3}{50}}a$ 
	& $-\sqrt{\frac{1}{10}}a$ & $-\frac{6}{5}a$ 
	& $-\frac{6}{5}a$ 
\\ \hline \hline 
\end{tabular} 
\label{table:coeff_rge}
\end{center}
\end{table}

The minimal $E_6$ model which have three generations of {\bf 27} 
and a pair ${\bf 2} + {\bf \ov{2}}$ depends in principle on the 
three cases; $H'$ has the same quantum number as $L$ or $H_d$ of 
{\bf 27}, or $\ov{H_u}$ of ${\bf\ov{27}}$. 
In the following we represent the hypercharge and the U(1)$'$ 
quantum numbers of the additional pair as $(-1/2,a)$ for $H'$ 
and $(+1/2,-a)$ for $\ov{H'}$, where the U(1)$'$ quantum number 
$a$ in each $Z'$ model follows the same normalization in 
Table~\ref{table_u1charge}. 

In the minimal model, 
the following eight scalar-doublets can develop VEV to 
give the mass terms $m_Z^2$ and $m_{ZZ'}^2$ in Eq.~(\ref{eq:l_gauge}): 
three generations of $H_u,  H_d$, 
and an extra pair, $H'$ and $\ov{H'}$. 
We express their VEVs as follows:
\bea
	\disp{\sum_{i=1}^3\langle H_u^i \rangle^2 = \frac{v_u^2}{2},}
~~
	\disp{\sum_{i=1}^3\langle H_d^i \rangle^2 = \frac{v_d^2}{2}, }
~~
	\disp{\langle H' \rangle^2 = \frac{v_{H'}^2}{2}, }
~~
	\disp{\langle \ov{H'} \rangle^2 = \frac{v_{\ov{H'}}^2}{2}}, 
\eea
where $i$ is the generation index. 
The sum of these VEVs gives the observed $\mu$-decay constant: 
$v_u^2 + v_d^2 + v_{H'}^2 + v_{\ov{H'}}^2 \equiv 
v^2 = \frac{1}{\sqrt{2}G_F} \approx (246~{\rm GeV})^2$. 
By further introducing the notation
\bea
\tan \beta = \frac{v_u}{v_d}, 
~~
x^2 = \frac{v_{H'}^2 + v_{\ov{H'}}^2}{v^2}, 
\eea
we can express $\zeta$ in Eq.~(\ref{eq:zeta}) as~\cite{chu} 
\bea
\zeta &=& 2 \biggl\{
	-Q_E^{H_u}(1-x^2)\sin^2\beta +Q_E^{H_d}(1-x^2)\cos^2\beta 
	+Q_E^{H'}x^2
	\biggr\} 
	- \delta. 
\eea
Because $H'$ and $\ov{H'}$ are taken from {\bf 27} + {$\bf\ov{27}$}, 
the ${\rm U(1)'}$ charge of $H'$, $Q_E^{H'}$, is identified with 
that of $L$, $H_d$ or $\ov{H_u}$. 

Let us remind the reader that, in the $\chi$ model, three generations 
of the matter fields {\bf 16} and  a pair of Higgs doublets make 
the model anomaly free. 
In this case, $\zeta$ is found to be independent of $\tan\beta$: 
\bea
\zeta &=& 2 Q_E^{H_d} - \delta. 
\eea
\begin{table}[t]
\begin{center}
\tcaption{Predictions for $g_E$ and $\delta$ at $\mu=m_{Z_1}$ 
	in the minimal models 
	and the $\eta_{\rm BKM}$ model~\cite{eta_model}. 
	The ${\rm U(1)}_Y$ gauge coupling $g_Y$ is fixed as 
	$g_Y = 0.36$. }
\begin{tabular}{ccccc}
\\ \hline \hline 
model & $a$ & $g_E$ & $g_E/g_Y$ & $\delta$ \\
\hline 
$\chi(16)$ & $\hph 3$ & $\hph 0.353$& $0.989$ & $\hph 0.066$ \\
	   & $-2$ & $\hph 0.361$ & $1.010$ &$-0.044$ \\
\hline
$\chi(27)$ & $\hph 3$ & $\hph 0.353$ & $0.989$ &$\hph 0.066$ \\
	   & $-2$ & $\hph 0.361$ & $1.010$ & $-0.044$ \\
\hline
$\psi$ & $\hph 1$ & $\hph 0.364$ & $1.020$ &$\hph 0.028$ \\
	& $\hph 2$& $\hph 0.356$ & $0.999$ &$\hph 0.056$ \\
	&$-2$& $\hph 0.356$ & $0.999$ & $-0.056$ \\
\hline
$\eta$ &$\hph 1/6$ & $\hph 0.366$ & $1.025$ &$\hph 0.018$ \\
	&$-2/3$ & $\hph 0.351$ & $0.982$ &$-0.071$ \\
\hline
$\nu$ & $\hph \sqrt{1/6}$ & $\hph 0.361$& $1.010$ & $\hph 0.044$ \\
	&$-\sqrt{3/8}$ & $\hph 0.353$ & $0.989$ & $-0.066$ 
\\ \hline 
$\eta_{\rm BKM}$~\cite{eta_model} 
	& ---  & $\hph 0.308$ & $0.862$ &$\hph 0.286$ 
\\ \hline \hline 
\end{tabular}
\label{table:ge_delta}
\end{center}
\end{table}
We can now examine the kinetic mixing parameter $\delta$ in each model. 
The boundary condition of $\delta$ at the GUT scale is $\delta = 0$. 
The non-zero kinetic mixing term can arise at low-energy 
scale through the following RGEs: 
\bsub
\bea
\frac{d}{dt} \alpha_i &=& \frac{1}{2\pi}b_i \alpha_i^2,  
\label{eq:rgea}
\\
\frac{d}{dt} \alpha_4 &=& \frac{1}{2\pi}
( b_E + 2 b_{1E} \delta + b_1 \delta^2 ) \alpha_4^2 ,
\label{eq:rgeb}\\
\frac{d}{dt} \delta &=& \frac{1}{2\pi}
( b_{1E} + b_1 \delta ) \alpha_1, 
\label{eq:rgec}
\eea
\label{eq:rge}
\esub
where $i=1,2,3$ and $t=\ln \mu$. 
We define $\alpha_1$ and $\alpha_4$ as 
\bea
\alpha_1 \equiv \frac{5}{3}\frac{g_Y^2}{4\pi}, 
~~~~~~~
\alpha_4 \equiv \frac{5}{3}\frac{g_E^2}{4\pi}. 
\eea
The coefficients of the $\beta$-functions for $\alpha_1, 
\alpha_4$ and $\delta$ are: 
\bea
b_1 = \frac{3}{5} {\rm Tr} (Y^2), 
~~~~
b_E = \frac{3}{5} {\rm Tr} (Q_E^2), 
~~~~
b_{1E} = \frac{3}{5} {\rm Tr} (Y Q_E). 
\eea
From Eq.~(\ref{eq:rgec}), we can clearly see that the non-zero $\delta$ 
is generated at the weak scale if $b_{1E} \neq 0$ holds. 
In Table~\ref{table:coeff_rge}, we list $b_1, b_E$ and $b_{1E}$ 
in the minimal $\chi,\psi,\eta$ and $\nu$ models. 
As explained above, the $\chi(16)$ model has three generations 
of {\bf 16}, and the $\chi(27)$ model has three generations of 
{\bf 27}. 
We can see from Table~\ref{table:coeff_rge} that the 
magnitudes of the differences $b_1 - b_2$ and $b_2 - b_3$ are 
common among all the models including the minimal supersymmetric 
SM (MSSM). 
This guarantees the gauge coupling unification 
at $\mu = m_{GUT} \simeq 10^{16}~{\rm GeV}$. 
It is straightforward to obtain $g_E(m_{Z_1})$ and 
$\delta(m_{Z_1})$ for each model. 
The analytical solutions of Eqs.~(\ref{eq:rgea})$\sim$
(\ref{eq:rgec}) are given in Ref.~\ref{ref:chu}. 
In our calculation, 
$\alpha_3(m_{Z_1}) = 0.118$ and 
$\alpha(m_{Z_1}) = e^2(m_{Z_1})/4\pi = 1/128$ 
are used as example. 
These numbers give $g_Y(m_{Z_1}) = 0.357$. 
We summarize the predictions for $g_E$ 
and $\delta$ at $\mu = m_{Z_1}$ in the 
minimal $E_6$ models in Table~\ref{table:ge_delta}. 
In all the minimal models, the ratio $g_E/g_Y$ is approximately 
unity and $|\delta|$ is smaller than about 0.07. 
Some further extra fields, therefore, may be needed to give 
$\delta=0.2$ which leads to the ``minimal $\Del \chi^2$'' when 
$\beta_E=-\pi/4$, which we found in Fig.~\ref{chisq_distribution}. 
We also show the result of the quasi leptophobic $\eta$ 
model ($\eta_{\rm BKM}$) proposed by Babu \etal~\cite{eta_model} 
in the same table. 
The $\eta_{\rm BKM}$ has, besides three generation of {\bf 27}, 
two pairs of $\bf{2 + \ov{2}}$ from ${\bf 78}$ and a pair of 
$\bf{3 + \ov{3}}$ from $\bf{27 + \ov{27}}$ in order to 
achieve the leptophobity ($\delta\sim 1/3)$ at the weak scale 
through the quantum corrections.  
We find that the $\eta_{\rm BKM}$ model predicts 
$g_E/g_Y \sim 0.86$ and $\delta \sim 0.29$, which is rather 
close to the leptophobity, $\delta=1/3$. 
\begin{table}[t]
\caption{Predictions for the effective $Z$-$Z'$ mixing parameter 
	$\zeta$ in the minimal $\chi,\psi,\eta$ and $\nu$ models 
	for $x^2 = 0$ and $0.5$, and $\tan\beta = 2$ and $30$. 
	}
\begin{center}
\begin{tabular}{c|c|c|c|c|c} \hline \hline
\multicolumn{2}{c|}{}& \multicolumn{2}{c|}{$x^2 = 0$} 
	& \multicolumn{2}{c}{$x^2 = 0.5$} 
\\ \cline{3-6}
\multicolumn{2}{c|}{}& \multicolumn{2}{c|}{$\tan\beta$}
	& \multicolumn{2}{c}{$\tan\beta$} 
\\ \hline
& $a$ & 2 & 30 & 2 & 30 
\\ \hline
$\chi$ & $\hph 3$ & \multicolumn{2}{c}{$-0.88$} 
	& \multicolumn{2}{c}{$0.14$} 
\\
     & $-2$ & \multicolumn{4}{c}{$-0.77$}
\\ \hline
$\psi$ &$\hph 1$ & $\hph 0.60$ & $\hph 1.02$ & $\hph 0.55$ 
	& $\hph 0.76$
\\
       &$\hph 2$ & $\hph 0.58$ & $\hph 1.00$ & $\hph 0.79$ 
	& $\hph 1.00$
\\
       & $-2$& $\hph 0.69$ & $\hph 1.11$ & $-0.16$ 
	& $\hph 0.06$
\\ \hline
$\eta$ & $\hph 1/6$ & $-1.02$ & $-1.35$ & $-0.35$ & $-0.52$ 
\\ 
& $-2/3$ & $-0.93$ & $-1.26$ & $-1.11$ & $-1.26$ 
\\ \hline
$\nu$ & $\hph \sqrt{1/6}$ & $\hph 0.36$ & $\hph 0.77$ & $\hph 0.57$ 
	& $\hph 0.77$ 
\\ 
 & $-\sqrt{3/8}$ & $\hph 0.47$ & $\hph 0.88$ & $-0.34$ 
	& $-0.14$ 
\\ \hline \hline
\end{tabular}
\end{center}
\label{table:zetasummary}
\end{table}

Next we estimate the parameter $\zeta$ for several 
sets of $\tan\beta$ and $x$. 
In Table~\ref{table:zetasummary}, we show the predictions for 
$\zeta$ in the minimal $\chi,\psi,\eta$ and $\nu$ models. 
The results are shown for $\tan\beta = 2$ and $30$, and 
$x^2 = 0$ and $0.5$. 
We find from the table that the parameter $\zeta$ is 
in the range $|\zeta| \lsim 1.35$. 
It is shown in Fig.~\ref{mass_95cl}
that $m_{Z_2}g_Y/g_E$ is approximately 
independent of $g_E/g_Y$.  
Actually, we find in Table~\ref{table:ge_delta} and 
Table~\ref{table:zetasummary} that 
the predicted $|\delta|$ is smaller than about 0.1 
and $g_E/g_Y$ is quite close to unity in all the minimal models.  
We can, therefore, read off from Fig.~\ref{mass_95cl} 
the lower bound of $m_{Z_2}$ in the minimal models at $g_E = g_Y$. 
In Table~\ref{table:mass95_zeta}, we summarize the 95\% CL lower 
$m_{Z_2}$ bound for the minimal $\chi,\psi,\eta$ and $\nu$ models 
which correspond to the predicted $\zeta$ 
in Table~\ref{table:zetasummary}. 
\begin{table}[t]
\caption{Summary of the 95\% CL lower bound of $m_{Z_2}$ 
	(GeV) which corresponds to the predicted $\zeta$ in 
	Table~\ref{table:zetasummary}. 
	}
\begin{center}
\begin{tabular}{c|c|r|r|r|r} \hline \hline
\multicolumn{2}{c|}{}& \multicolumn{2}{c|}{$x^2 = 0$} 
	& \multicolumn{2}{c}{$x^2 = 0.5$} 
\\ \cline{3-6}
\multicolumn{2}{c|}{}& \multicolumn{2}{c|}{$\tan\beta$}
	& \multicolumn{2}{c}{$\tan\beta$} 
\\ \hline
& $a$ &$~~~~~~2$ & $~~~~30$ & $~~~~~~2$  & $~~~~30$
\\ \hline
$\chi$ & $\hph 3$ & \multicolumn{2}{c}{$1490$} 
	& \multicolumn{2}{c}{$620$} 
\\
     & $-2$ & \multicolumn{4}{c}{$1380$}
\\ \hline
$\psi$ &$\hph 1$ & $1270$ & $1780$ & $1200$ & $1470$
\\
       &$\hph 2$ & $1250$ & $1760$ & $1510$ & $1760$
\\
       & $-2$& $1380$ & $1890$ & $540$ & $370$
\\ \hline
$\eta$ & $+1/6$ & $1440$ & $1840$ & $660$ & $860$ 
\\ 
& $-2/3$ & $1330$ & $1730$ & $1550$ & $1730$
\\ \hline
$\nu$ & $+\sqrt{1/6}$ & $1030$ & $1600$ & $1340$ & $1600$ 
\\ 
& $-\sqrt{3/8}$ & $1200$ & $1730$ & $880$ & $580$ 
\\ \hline \hline
\end{tabular}
\end{center}
\label{table:mass95_zeta}
\end{table}
Most of the lower mass bounds in Table~\ref{table:mass95_zeta} 
exceed 1 TeV. 
The $Z_2$ boson with $m_{Z_2} \sim O(1~{\rm TeV})$ should be 
explored at the future collider such as LHC. 
The discovery limit of the $Z'$ boson in the $E_6$ models at LHC 
is expected as (in unit of $\gev$)~\cite{cvetic_bound}
\bea
\begin{array}{cccc}\hline 
\chi & \psi & \eta & \nu \\ \hline 
3040 & 2910 & 2980 & *** \\ \hline 
\end{array}
\eea
All the lower bounds of $m_{Z_2}$ listed in 
Table~\ref{table:mass95_zeta} are smaller than 2 TeV 
and they are, therefore, in the detectable range of LHC. 
But, it should be noticed that most of them 
($1~{\rm TeV} \lsim m_{Z_2}$) may require the fine-tuning to stabilize 
the electroweak scale against the ${\rm U(1)'}$ scale~\cite{drees}. 

%
%
%
%
\section{Summary}
In this review article, 
we have studied constraints on $Z'$ bosons in the SUSY 
$E_6$ models. 
Four $Z'$ models --- the $\chi,\psi,\eta$ and $\nu$ models 
are studied in detail. 
The presence of the $Z'$ boson affects the electroweak processes 
through the effective $Z$-$Z'$ mass mixing angle $\xibar$, 
a tree level contribution $T_{\rm new}$ and the contact term 
$\contact$, where the latter two parameters are positive definite 
quantities. 
The $Z$-pole, $\mw$ and LENC data constrain ($T_{\rm new}, 
\xibar$), $T_{\rm new}$ and $\contact$, respectively. 
From the updated electroweak data, we find that three $Z'$ models 
($\chi,\eta,\nu$) improve the fit over the SM where the total 
$\chi^2$ decrease about five units, owing to the excellent fit 
mainly to the improved 
data of parity violation in cesium atom which is expressed by 
the weak charge $Q_W(^{133}_{55}Cs)$. 
The more than 2-$\sigma$ deviation of $Q_W(^{133}_{55}Cs)$ from 
the SM prediction could be explained in these three $Z'$ models. 
Due to its parity conserving property of the U(1)$'$ charge 
assignment on the SM matter fields, the $\psi$ model does not 
improve the fit to the $Q_W(^{133}_{55}Cs)$ data. 
The impact of the kinetic mixing $(\delta \neq 0)$ on the fit 
is also examined on the $(\beta_E,\delta)$ plane. 
The $Z'$ model with ($\beta_E, \delta)=(-\pi/4, 0.2$) shows 
the most excellent fit to the data among the SUSY $E_6$ models 
where the total $\chi^2$ 
decreases by about seven units as compared to the SM best fit. 
The 95\% CL lower mass bound of the heavier mass eigenstate 
$Z_2$ is shown as a function of the effective $Z$-$Z'$ mixing 
parameter $\zeta$ together with the result of direct search 
experiment. 
By assuming the minimal particle content of the $E_6$ model,  
we have found the theoretical predictions for $\zeta$. 
It is shown that the $E_6$ models with minimal particle content 
which is consistent with the gauge coupling unification predict 
the non-zero kinetic mixing term $\delta$ and the effective mixing 
parameter $\zeta$ of order one. 
The present electroweak experiments lead to the lower mass bound 
of order 1 TeV or larger for those models. 

%
%
%
%
\nonumsection{Acknowledgements}

The author would like to thank K. Hagiwara and Y. Umeda for 
fruitful collaborations which this report is based upon. 
He is also grateful to R. Barbieri for reading manuscript. 


%
%
%
%
%
%
\nonumsection{References}


\begin{thebibliography}{99}
\bibitem{hewett_rizzo}
	\label{ref:hewett_rizzo}
	J. Hewett and T. Rizzo, \PR{183}{1989}{193}.
\bibitem{radiative_u1prime}
	\label{ref:radiative_u1prime}
	M. Cveti\v{c} \etal, \PRD{56}{1997}{2861}; 
	P. Langacker and J. Wang, \PRD{58}{1998}{115010}. 
\bibitem{chu}
	\label{ref:chu}
	G.C. Cho, K. Hagiwara and Y. Umeda, 
	\NPB{531}{1998}{65}; {\bf B555} (1999) 651 (E). 
\bibitem{uch}
	\label{ref:uch}
	Y. Umeda, G.C. Cho and K. Hagiwara, \PRD{58}{1998}{115008}. 
\bibitem{zprime_old}
	G. Altarelli \etal, \PLB{263}{1991}{459}; 
	F. del Aguila, W. Hollik, J.M. Moreno and M. Quiros, 
	\NPB{372}{1992}{3}; 
	P. Langacker and M. Luo, \PRD{45}{1992}{278}; 
	G. Altarelli \etal, \PLB{318}{1993}{139};  
	P. Langacker, in {\it Precision Tests of the Standard 
	Electroweak Model}, (ed) P. Langacker, 
	World Scientific, (1995) 883; 
	T. Gherghetta, T.A. Kaeding and G.L. Kane, 
	\PRD{57}{1998}{3178}. 
\bibitem{chm}
	\label{ref:chm}
	G.C. Cho, K. Hagiwara and S. Matsumoto, 
	\EPJC{5}{1998}{155}. 
\bibitem{zprime_lep}
	L3 Collaboration, \PLB{306}{1993}{187}; 
	ALEPH Collaboration, \ZPC{62}{1994}{539}. 
\bibitem{erler_langacker}
	J. Erler and P. Langacker, \PLB{456}{1999}{68}. 
\bibitem{apv_data}
	S.C. Bannett and C.E. Wieman, 
	\PRL{82}{1999}{2484}.
\bibitem{apv_zprime}
	R. Casalbuoni \etal, \PLB{460}{1999}{135}; 
	J.L. Rosner, \PRD{61}{2000}{016006}; 
	J. Erler and P. Langacker, \PRL{84}{2000}{212}. 
\bibitem{holdom}
	B. Holdom, \PLB{166}{1986}{196}. 
\bibitem{eta_model}
	\label{ref:eta_model}
	K.S. Babu, C. Kolda and J. March-Russell, 
	\PRD{54}{1996}{4635}.
\bibitem{general_zzmixing}
	\label{ref:general_zzmixing}
	K.S. Babu, C. Kolda and J.March-Russell, 
	\PRD{57}{1998}{6788}.
\bibitem{rizzo_umeda}
	\label{ref:rizzo_umeda}
	T. Rizzo, \PRD{59}{1999}{015020};  
	Y. Umeda, Ph.D. thesis, Hokkaido University, Japan (1999). 
\bibitem{peskin_takeuchi}
    M.E. Peskin and T. Takeuchi, 
    \PRL{65}{1990}{964}; \PRD{46}{1992}{381}.
\bibitem{PDG98}
	\label{ref:PDG98}
	Review of Particle Physics, 
	C. Casao \etal, \EPJC{3}{1998}{1}
\bibitem{eta_ellis}
	J. Ellis, K. Enqvist, D.V. Nanopoulos and F. Zwirner, 
	\MPLA{1}{1986}{57}; \NPB{276}{1986}{14}. 
\bibitem{nu_model}
	L.E. Ib\'{a}\~{n}ez and J. Mas, \NPB{286}{1987}{107}; 
	T. Matsuoka, H. Mino, D. Suematsu and 
	S. Watanebe, \PTP{76}{1986}{915}. 
\bibitem{see-saw} 
	T. Yanagida, in {\it Proceedings of the Workshop on the 
	Unified Theory and the Baryon Number in the Universe}, 
	(ed) O. Sawada and A. Sugamoto, KEK (1979); 
	M. Gell-Man, P. Ramond and R. Slansky, 
	in {\it Supergravity}, (ed) P.von Nienvenhuzen and 
	D.Z. Freedman, North Holland (1979). 
\bibitem{hhkm}
	\label{ref:hhkm}
	    K. Hagiwara, D. Haidt, C.S. Kim and S. Matsumoto, 
	    \ZPC{64}{1994}{559}; {\bf C68} (1995) 352 (E).
\bibitem{holdom2}
	\label{ref:holdom2}
        B. Holdom, \PLB{259}{1991}{329}. 
\bibitem{lepewwg98}
	\label{ref:lepewwg98}
	The LEP Collaborations ALEPH, DELPHI, L3, OPAL, 
	the LEP Electroweak Working Group and the SLD Heavy 
	Flavor Group, CERN-EP/99-15. 
\bibitem{wboson_moriond} 
	\label{ref:wboson_moriond} 
	I.~Riu, talk given at the XXXIVth Rencontres de Moriond, 
	March 13-20, 1999. 
\bibitem{slac}
	C.Y. Prescott \etal, \PLB{84}{1979}{524}. 
\bibitem{cern}
	A. Argento \etal, \PLB{120}{1983}{245}. 
\bibitem{bates}
        P.A. Souder \etal, \PRL{65}{1990}{694}.
\bibitem{mainz}
        W. Heil \etal, \NPB{327}{1989}{1}.
\bibitem{fh}
        G.L. Fogli and D. Haidt, \ZPC{40}{1988}{379}
\bibitem{ccfr}
        K. McFarland \etal, \EPJC{1}{1998}{509}.
\bibitem{charm-II} 
	CHARM-II Collaboration, \PLB{281}{1992}{159}. 
\bibitem{EJ} 
	\label{ref:EJ} 
	S. Eidelman and F. Jegerlehner, \ZPC{67}{95}{585}.
\bibitem{hhm}
	\label{ref:hhm}
	K. Hagiwara, D. Haidt and S. Matsumoto, 
	\EPJC{2}{1998}{95}. 
\bibitem{jekim}
	J.E. Kim, P. Langacker, M. Levine and H.H. Williams, 
	Rev. Mod. Phys. {\bf 53} (1981) 211.
\bibitem{higgs_moriond}
	M. Felcini, talk given at the XXXIVth Rencontres 
	de Moriond, March 13-20, 1999. 
\bibitem{kane}
	G.L. Kane, C. Kolda and J.D. Wells, \PRL{70}{1993}{2686}; 
	D. Comelli and C. Verzegnassi, \PLB{303}{1993}{277}.
\bibitem{direct_search}
	CDF Collaboration, \PRL{79}{1997}{2192}. 
\bibitem{drees}
	M. Drees, N.K. Falck and M. Gl\"{u}ck, 
	\PLB{167}{1986}{187}.
\bibitem{dienes}
	K. Dienes, Phys. Rep. {\bf 287} (1997) 447. 
\bibitem{cvetic_bound}
	M. Cveti\v{c} and S. Godfrey, summary of the Working 
	Subgroup on Extra Gauge Bosons of the DPF long-range 
	planning study, 
	in {\it Electro-weak Symmetry Breaking and Beyond the 
	Standard Model}, (ed). T. Barklow \etal, World Scientific 
	(1995) (hep-ph/9504216). 
\end{thebibliography}
\end{document}